\documentclass[aps,prl,twocolumn,amsmath,amssymbs,superscriptaddress]{revtex4-1}

\usepackage{graphicx}
\usepackage{amsmath,amsfonts,amssymb}
\usepackage{epsfig}
\usepackage{wrapfig}
\usepackage{bbm}
\usepackage[usenames]{color}
\usepackage{array}
\usepackage{times}

\usepackage{xcolor}

\usepackage{hyperref}

\newcommand{\eps}{\varepsilon}

\begin{document}

\title{Multipole higher-order topology in a multimode lattice
}

\author{Maxim Mazanov}
\thanks{These three authors contributed equally}
\affiliation{School of Physics and Engineering, ITMO University, Saint Petersburg 197101, Russia}

\author{Anton S. Kupriianov}
\thanks{These three authors contributed equally}
\affiliation{State Key Laboratory of Integrated Optoelectronics, College of Electronic Science and Engineering, International Center of Future Science, Jilin University, 2699 Qianjin Street, Changchun 130012, China} 

\author{Roman S. Savelev}
\thanks{These three authors contributed equally}
\affiliation{School of Physics and Engineering, ITMO University, Saint Petersburg 197101, Russia}

\author{Zuxian~He}
\affiliation{State Key Laboratory of Integrated Optoelectronics, College of Electronic Science and Engineering, International Center of Future Science, Jilin University, 2699 Qianjin Street, Changchun 130012, China}

\author{Maxim A. Gorlach}
\email{m.gorlach@metalab.ifmo.ru}
\affiliation{School of Physics and Engineering, ITMO University, Saint Petersburg 197101, Russia}

\begin{abstract}
The concepts of topology have a profound impact on physics research spanning the fields of condensed matter, photonics and acoustics and predicting topological states that provide unprecedented versatility in routing and control of waves of various nature. Higher-order topological insulators further expand this plethora of possibilities towards extended range of structure dimensionalities. 
Here, we put forward a novel class of two-dimensional multipolar higher-order topological insulators that arise due to the interference of the degenerate modes of the individual meta-atoms generalizing the mechanism of spin-orbit coupling in condensed matter systems. We prove that this model features disorder-robust corner modes and cannot be reduced to the known crystalline topological phases or conventional quadrupole insulators, providing the first example of multipolar topology in a $C_3$-symmetric lattice featuring quantized octupole moment. 
The multimode nature of the lattice gives rise to flat bands and corner states with extreme localization enabling coherent control of the topological modes. We support our predictions by assembling the designed structure, observing multipolar topological corner states and experimentally demonstrating their coherent control.
\end{abstract}

\date{\today}

\maketitle



\section{Introduction}

Since the discovery of quantum Hall effect, topological phases of matter have uncovered new avenue of fundamental physics which predicts localized states protected against disorder by the global symmetries of the system~\cite{Lu2014Nov,Lu2016,Ozawa_RMP_2019}. Rich plethora of topological phenomena includes valley Hall effect~\cite{Noh2018Feb,Zhong2020Jul,Mak2014Jun,Dong2017Mar,Lundt2019Aug}, nonlinear~\cite{Smirnova2020} and non-Hermitian~\cite{Bergholtz2021,Leykam2017} topological systems, topological lasers~\cite{Harari2018,Bandres2018,Zeng2020} 
and, since recently, higher-order topological states~\cite{Benalcazar_2017_Science,Benalcazar_2017_PRB,Schindler2018,Xie_HOTI_review_2021}. The latter are characterized by the localization in at least two dimensions less than the dimensionality of the system. 

Introduction of spin degrees of freedom and spin-orbit interactions originally proposed in condensed-matter systems~\cite{Kane_Mele_2005,Kane2005Nov} 
has naturally lead to the discovery of new topological classes generalizing the existing ones, such as 
quantum 
spin Hall effect~\cite{Fu_Kane2006,Wu_Hu,Barik2018Feb,Xie2020Jul,Slobozhanyuk2017Feb}, or spinful versions of higher-order topological structures~\cite{Schindler_2019,GladsteinGladstone2022Jan,Zhang2020Jun,Zhu2021Jul,Chen2021Mar,Yang2021Apr}. 
However, condensed matter systems pose stringent limitations on possible values of the spin and its couplings to the orbital motion. In contrast, photonics promises a viable alternative to realize arbitrary pseudospin degrees of freedom exploiting accidental degeneracy of the modes in the meta-atoms. Since the number of the degenerate modes, their rotational symmetry and their couplings can be flexibly tailored, this provides a direct access to spinful systems with exotic types of spin-orbit coupling beyond those available in condensed matter which in turn may enable novel types of topology.  


As a potential platform to probe such models one may use coupled waveguides~\cite{Rechtsman2013Apr,Noh2018Jul,Zilberberg2018Jan,Caceres-Aravena2022Jun,Guzman-Silva2021Aug}, polaritonic micropillar lattices~\cite{Bloch2017Oct} and photonic crystals~\cite{Wu_PRL_2007,Xie2020Jul}. Regardless of the details of experimental implementation, our strategy to tailor the topological properties remains qualitatively the same. Due to the distinct rotational symmetry of the degenerate modes further referred as orbitals, the interference of their near fields allows us to modulate the effective coupling between the elements creating analogs of spin-orbit interaction in solids and even go beyond that. 
As a consequence, simple lattices enable novel topological phases featuring disorder-robust modes and lifting strong requirements to the lattice symmetry with topological protection rooted in the degeneracy of on-site orbital modes.

\begin{figure*}
	\centering
	\includegraphics[width=0.98\textwidth]{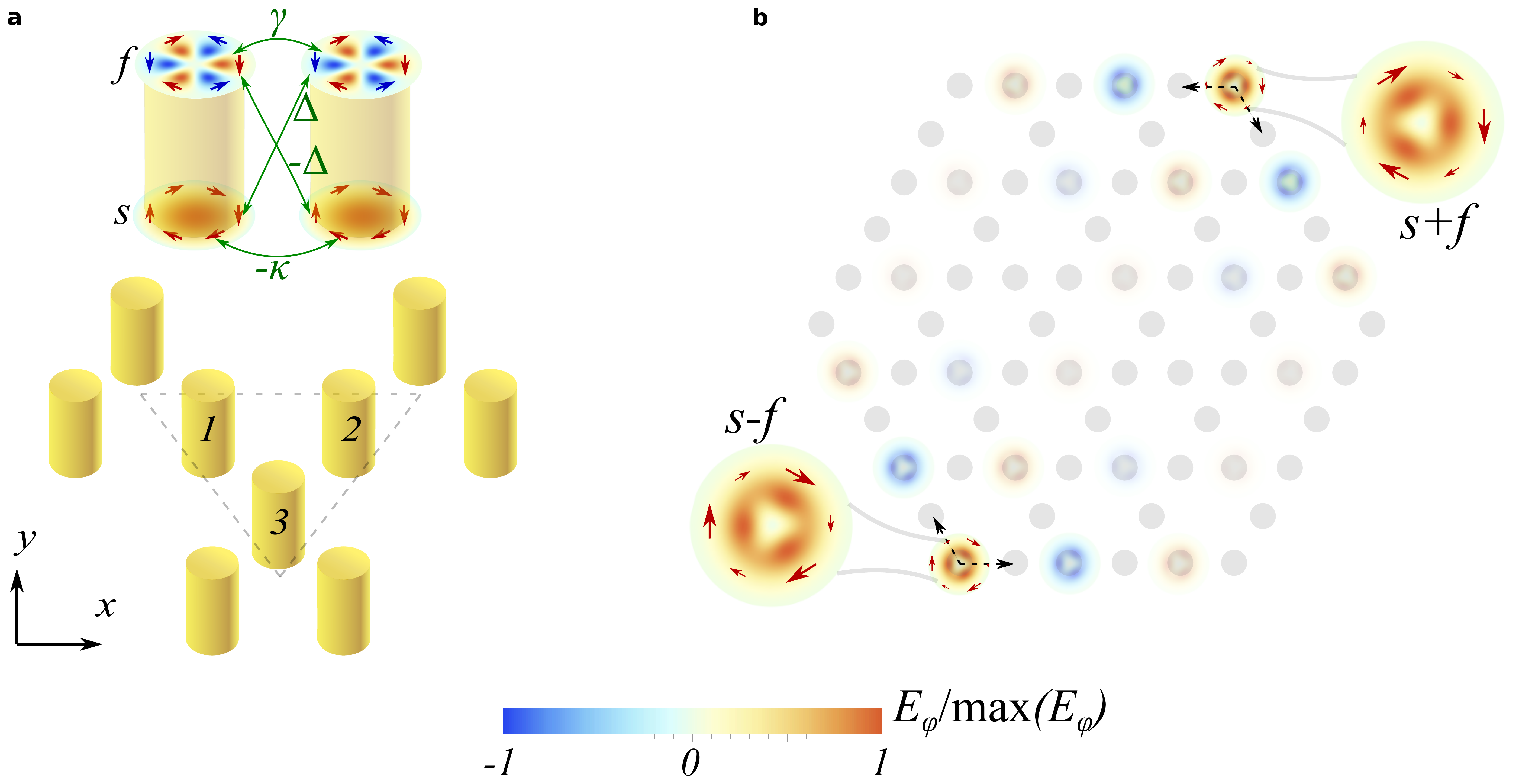}
	\caption{
		\textbf{Design of the system and topological corner states.} 
		\textbf{a}, Sketch of the proposed waveguide kagome lattice with $s$-like TE$_{0n}$ and $f$-like HE$_{3m}$ orbital modes with arbitrary radial indices $m,n$ 
		at each site. Different couplings between the orbitals $-\kappa, \gamma, \pm\Delta$ are shown schematically by green arrows. 
		\textbf{b}, Structure of the two corner states localized at the opposite corners. Black dashed arrows mark the directions from the corner sites to their two nearest neighbours in which the effective coupling vanishes provided $\kappa=\gamma=\Delta$. 
		In both panels, the amplitude of the azimuthal component of the electric field $E_\varphi$ responsible for the couplings is shown schematically by the arrows and color-coded in 
		the amplitude profiles. 
		The qualitative picture remains the same if TM$_{0n}$ and HE$_{3m}$ orbital modes are chosen instead. 
	}
	\label{Fig1}
\end{figure*}

Here, we propose to exploit multi-mode meta-atoms in the lattices with a relatively simple primitive cell, where inter-mode coupling opens the topological gap and generates multipole higher-order topological insulator (HOTI) phase.  
To exemplify this concept, we consider an undistorted photonic kagome lattice composed of evanescently coupled waveguides. The waveguides are designed to have doubly degenerate modes in the frequency range of interest. The dependence of the near field on azimuthal angle is captured by the factor $e^{in\varphi}$, where $n$ is equal either to 0 or to $\pm 3$, which allows us to classify the modes as $s$ and $f$ orbitals, respectively (Fig.~\ref{Fig1}a). Deriving the theoretical coupled-mode description of the designed lattice, we prove the emergence of multipole higher-order topology in our system and confirm our prediction by the experimental observation of topological corner states. As we highlight, this system cannot be reduced to the known crystalline topological insulators~\cite{Quantization_2019,Wu_Hu,Yves2017} or canonical quadrupole insulators~\cite{Benalcazar_2017_Science,Benalcazar_2017_PRB,Peterson2018,Photonic_quadrupole_2019,Hassan2019,Mele_2020,Hua_PRB_2020,Hua_LPR_2020} and thus provides a distinct type of higher-order topology in $C_3$-symmetric lattices mediated by the engineered effective spin-orbit coupling. 
Furthermore, as we prove, our two-dimensional model features quantized octupole moment, which hints towards octupolar origin of higher-order topology.

In addition to nontrivial topological properties and associated corner states, our model exhibits completely flat bands for the suitably designed coupling constants, a feature previously found in two dimensions only in the presence of external magnetic fields~\cite{Vidal_PRL}. 


\begin{figure*}
	\centering
	\includegraphics[width=0.98\textwidth]{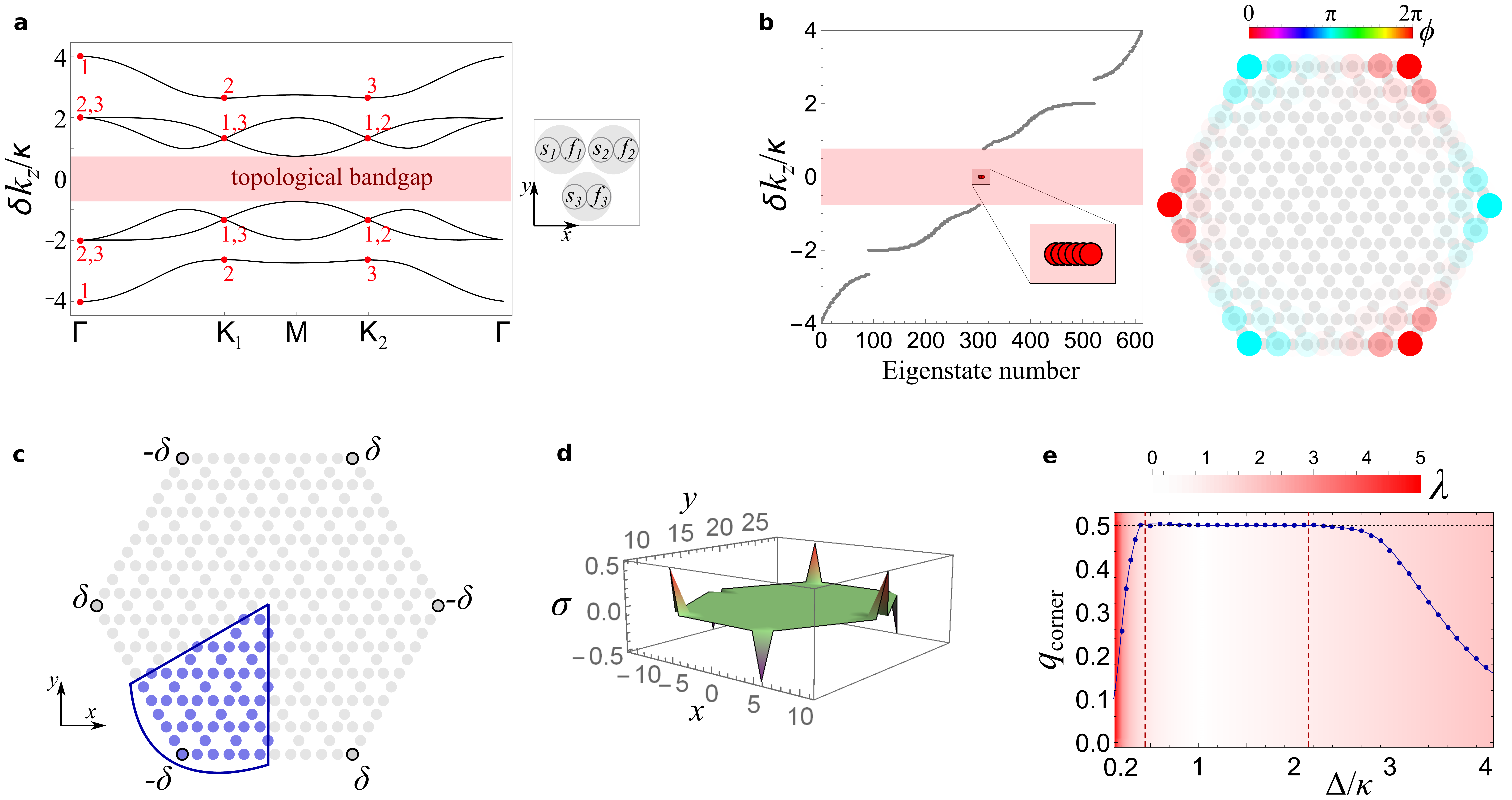}
	\caption{
		\textbf{Band structure and corner charges.} \textbf{a}, Band diagram of the structure for $\kappa=\gamma, \, \Delta=0.5 \kappa$. 
		The numbers at the curves show the value of symmetry index $p$ encoding $C_3$ rotation eigenvalue at a given wave number. 
		\textbf{b}, Spectrum of the finite lattice for the same parameters as in \textbf{a}, with six topological corner states in the bandgap (left), and the corresponding profiles of the corner states. The relative phase $\phi$ between $s$ and $f$ modes is color-coded (right). 
		\textbf{c,d}, Structure of the finite sample used to calculate the distribution of the effective charge density $\sigma(x,y)$ for $\kappa=\gamma=\Delta$ and the calculated results. The on-site potential $\delta = \pm 0.001 \kappa$ is added at the corners of the structure to fix the signs of the corner charges, while the highlighted sector shows the integration area for the corner charge. \textbf{e}, Corner charge $q_\text{corner}$ calculated for varying inter-orbital coupling $\Delta$ while keeping $\kappa=\gamma$ constant. The color-coded background in~\textbf{e} encodes the localization length $\lambda$ of the corner state [see Eq.~\eqref{lambda_loc}]. Vertical red dashed lines delimit the domain where $\lambda \leq 1$. 
	}
	\label{Fig2}
\end{figure*}

\section{Results}

\textbf{Topological multimode lattice.} 
To reveal the nature of the predicted topological phase, we design $C_3$-symmetric photonic kagome lattice depicted in Fig.~\ref{Fig1}a, where each waveguide hosts $s$ and $f$ orbital modes having the same propagation constants $k_z$ at a given frequency. Note that the profiles of both modes are compatible with $C_3$ symmetry of kagome unit cell. When arranged into lattice, modes of the same type ($s$-$s$ and $f$-$f$) couple with the strengths $-\kappa$ and $\gamma$, where the sign and magnitude of the coupling constant are determined by the overlap integral of the respective modes' electric fields~\cite{SavelevGorlach_PRB}. 
However, as we show below, the crucial ingredient in the formation of a topological phase is inter-orbital coupling $\Delta$ (Fig.~\ref{Fig1}a).

As a finite realization of our model, we consider a hexagon-shaped structure (Fig.~\ref{Fig1}b) which is in contrast with the  previous triangular-shaped implementations of kagome lattices~\cite{Xue2019Feb,Ni2019Feb,Li2020Feb}. 
Moreover, while the topological bandgap in a single-mode kagome lattice is opened due to the geometrical deformations (breathing kagome geometry), here we exploit a different approach based on the inter-orbital coupling.

To grasp the physics behind hybrid topological corner states, it is instructive to consider the interference of the two orbitals comprising the corner mode in the special case $\kappa=\gamma=\Delta$. In such scenario, one may construct an $s+f$ superposition localized at the upper right corner of the hexagon (see Fig.~\ref{Fig1}b). As can be readily read off from Fig.~\ref{Fig1}a, the coupling of such superposition to the adjacent $s$ and $f$-modes is equal to $-\kappa+\Delta$ and $\gamma-\Delta$, respectively. In this limiting case, both of these couplings vanish. Given that the waveguides are coupled evanescently and the next-nearest-neighbor couplings are negligible, the constructed corner state remains fully decoupled from the lattice and therefore is maximally localized. Performing spatial inversion of $s+f$ superposition and keeping in mind different parity of $s$ and $f$ orbitals, we recover that $s-f$ is another maximally localized corner mode at the opposite corner of the lattice (Fig.~\ref{Fig1}b). The localized states at four  remaining corners are either $s+f$ or $s-f$ superpositions  recovered from the pair of modes above by $C_3$ rotation. This simple analysis suggests that the geometry in Fig.~\ref{Fig1}b gives rise to the corner states and mode interference is a crucial ingredient for their formation. 

Our qualitative reasoning allows us to capture one more important feature of the system. Even though the finite sample in Fig.~\ref{Fig1}b looks $C_6$-symmetric, the symmetry of the model is only $C_3$ because of the $f$ modes with odd parity under inversion. This symmetry manifests itself in the wave functions of the corner modes: different corners of the lattice host either $s+f$ or $s-f$ hybrid modes (Fig.~\ref{Fig1}b). These superpositions are converted one to another either via spatial inversion or by $\pi/3$ rotation.

It should be noted that each $f$-mode is doubly degenerate due to time-reversal symmetry.  However, symmetry of the lattice and evanescent nature of the couplings ensure that only one mode from this doublet enters the Hamiltonian, while another $f$ mode remains decoupled in the nearest-neighbor approximation and does not affect the topological properties of the system.

\textbf{Theoretical model.} 
The outlined qualitative picture with the corner mode occupying a single site is valid only for $\kappa=\gamma=\Delta$. Now we refine our treatment deriving the tight-binding model of the system in the general case. 

In the basis of the modes $\textbf{v}=(s_1,f_1, s_2,f_2, s_3,f_3)^T$ shown schematically in Fig.~\ref{Fig2}a, Bloch Hamiltonian reads (see Supplemental Materials for derivation): 
\begingroup
\begin{equation}
\label{Hamk}
\hat{H}(\textbf{k})
=
\begin{pmatrix}
	0 & \hat{\kappa}_- + \alpha^\dag \hat{\kappa}_+ & \hat{\kappa}_+ + \alpha^\dag \beta \hat{\kappa}_- \\
	\hat{\kappa}_+ + \alpha \hat{\kappa}_- & 0 & \hat{\kappa}_- + \beta \hat{\kappa}_+ \\
	\hat{\kappa}_- + \alpha \beta^\dag \hat{\kappa}_+ & \hat{\kappa}_+ + \beta^\dag \hat{\kappa}_- & 0 \\
\end{pmatrix}
,
\end{equation}
\endgroup
where $ \textbf{k} = (k_x,k_y) $ is the Bloch wave vector, $\alpha \equiv e^{i \textbf{k} \textbf{a}}, \, \beta \equiv e^{i \textbf{k} \textbf{b}}$, $ \mathbf{a} = (1,0)^T, \, \mathbf{b} = (1/2,\sqrt{3}/2)^T $ are the primitive lattice vectors, and the coupling matrices read $ \hat{\kappa}_\pm = \left(
\begin{array}{cccccc}
-\kappa & \pm \Delta \\
\mp \Delta & \gamma \\
\end{array}
\right) $, $\kappa$, $\gamma$ and $\Delta$ being positive integers that depend on the  specific implementation of the model. The eigenvalue equation takes the form $ \hat{H}(\textbf{k}) \textbf{v} = \delta k_z
\textbf{v}
$, where $ \delta k_z = k_z - k_{z0} $ is the difference between the propagation constants of a collective mode and any of the two degenerate modes in an isolated waveguide. Note that the inter-orbital coupling $\Delta$ is odd under inversion due to the different parity of $s$ and $f$ modes which allows us to draw an analogy between this mechanism and spin-orbit coupling. This analogy can be extended further by presenting the Hamiltonian of our model as two spin blocks coupled by the effective spin-orbit coupling proportional to~$\Delta$. However, in contrast to the conventional spin-orbit coupling introduced in e.g. Kane-Mele model~\cite{Kane_Mele_2005} or its variations for kagome metals~\cite{Ye2018Mar,Guo2009Sep}, 
the mechanism introduced here involves spin-flip process which facilitates the formation of topological phase rather than destroying it as further analyzed in Supplementary Materials. 


Inspecting the derived Hamiltonian Eq.~\eqref{Hamk}, we observe that it exhibits all symmetries of the model including $C_3$ rotations and generalized chiral symmetry (see Supplemental Materials). 
The band structure features six bands of hybrid modes (Fig.~\ref{Fig2}a), whereas nonzero inter-orbital interaction $\Delta$ opens the topological bandgap in the dispersion.

To characterize the topological properties of our model, we analyze the symmetry indicators for the bulk bands, adopting the classification of higher-order topological crystalline insulators with time-reversal and $C_n$ symmetries~\cite{Quantization_2019}. 
As the model possesses $C_3$ symmetry, its topological properties are captured by the two topological indices, $ [K^{(3)}_1] $ and $ [K^{(3)}_2] $, where $ [K^{(3)}_p] = \# K^{(3)}_p - \# \Gamma^{(3)}_p $, and $ \# \Pi^{(3)}_p $ is the number of the bands below the gap at a particular high-symmetry point $ \boldsymbol{\Pi}^{(3)} $ which have the eigenvalue $ \Pi^{(3)}_p = e^{2\pi i (p-1)/3} $ with respect to the $C_3$ rotation. 
These symmetry eigenvalues $\Pi^{(3)}_p$ labelled for brevity by the index $p$  are indicated for $\Gamma, \text{K}_1, \text{K}_2$ points in Fig.~\ref{Fig2}a. From the derived indices, we obtain trivial topological invariants $ [K^{(3)}_1] = 0 $ and $ [K^{(3)}_2] = 0 $, which means that our model features zero bulk polarization, $\textbf{P}=0$.

Even though bulk polarization vanishes, a finite hexagonal sample with nonzero inter-orbital coupling $\Delta$ (Fig.~\ref{Fig2}b) hosts six hybrid corner states composed of $s$ and $f$ orbital modes. The relative phases $\phi$ between $s$ and $f$ modes depicted in Fig.~\ref{Fig2}b fully agree with the qualitative reasoning for the special case $\kappa=\gamma=\Delta$ discussed above. 
The localization length of the corner states can be evaluated analytically, and in the units of the lattice constant reads (see Supplementary Materials for derivation): 
\begin{equation}
\label{lambda_loc}
\lambda 
= 
\ln^{-1}{\left| 2\Delta / \left(\Delta - \sqrt{\gamma \kappa}\right) - 1 \right|}
. 
\end{equation}
In particular, the localization is perfect ($\lambda = 0$) for a continuous range of coupling constants, $\Delta=\sqrt{\gamma \kappa}$, which includes the flat-band case $\kappa=\gamma=\Delta$.
Furthermore, the corner states are robust to the moderate values of disorder in the coupling constants as well as to the homogeneous detuning between the orbital modes which hints towards their topological nature. The latter feature allows us to observe the corner states in the entire range of frequencies where degeneracy of the modes is no longer perfect, but the detuning remains relatively small:
\begin{equation}\label{eq:DetuningLimit}
    \delta k \leq 2( \gamma + \kappa )\:.
\end{equation}

Thus, we recover the phase with robust topological corner states but vanishing bulk polarization. This points towards multipole topology of our model. 
However, to the best of our knowledge, no $C_3$-symmetric multipole topological insulator has been found to date.  
Therefore, to confirm our hypothesis, we calculate the effective charge density distribution $\sigma(x,y)$ when alternating infinitesimal potential $\pm \delta$ is added at the corners of the structure to fix the signs of the corner charges, as shown in Fig.~\ref{Fig2}c (see the Methods section). 
The calculated distribution depicted in Fig.~\ref{Fig2}d exhibits an alternating $C_3$-symmetric pattern matching the relative phases of $s$ and $f$ modes in Fig.~\ref{Fig2}b. 
Interestingly, this arrangement of the corner charge by no means can be reduced to the known models of crystalline topological insulators possessing either nonzero or zero bulk polarization~\cite{Quantization_2019,Benalcazar_2017_PRB,Benalcazar_2017_Science}. This fact, together with the disorder robustness of the corner modes, suggests a novel type of multipolar topology.

To highlight the role played by the inter-orbital coupling in the emergence of higher-order topology in our system, we evaluate the corner charge $q_\text{corner}$ and show that it is robustly pinned to the value $0.5$ for a wide range of $\Delta$ when, according to Eq.~\eqref{lambda_loc}, localization length is so small $\lambda \lesssim 1$ that the interaction of the corner states at the adjacent corners can be neglected (Fig.~\ref{Fig2}e).

To further clarify the type of higher-order topology in our system, we calculate numerically the real-space quadrupole~\cite{Li2020Oct,Agarwala2020Mar} and octupole moments for the finite lattices with various coupling ratios $\gamma/\kappa$, $\Delta/\kappa$ (Supplementary Materials). To compute the octupole moments, we generalize the results of modern theory of polarization~\cite{Resta1998Mar}. As we show, while the independent components of quadrupole moment $q_1, q_2$ are not quantized simultaneously, both independent components of octupole moment $o_1, o_2$ are half-quantized for a wide range of coupling constants. These calculations suggest that the octupole moment can be viewed as a relevant topological invariant for our system which distinguishes it from the known quadrupole topological phases~\cite{Benalcazar_2017_Science,Benalcazar_2017_PRB}. In such interpretation, our system provides the first example of \textit{octupolar topology in a two-dimensional system.}

\begin{figure*}[t]
	\centering
	\includegraphics[width=0.99\textwidth]{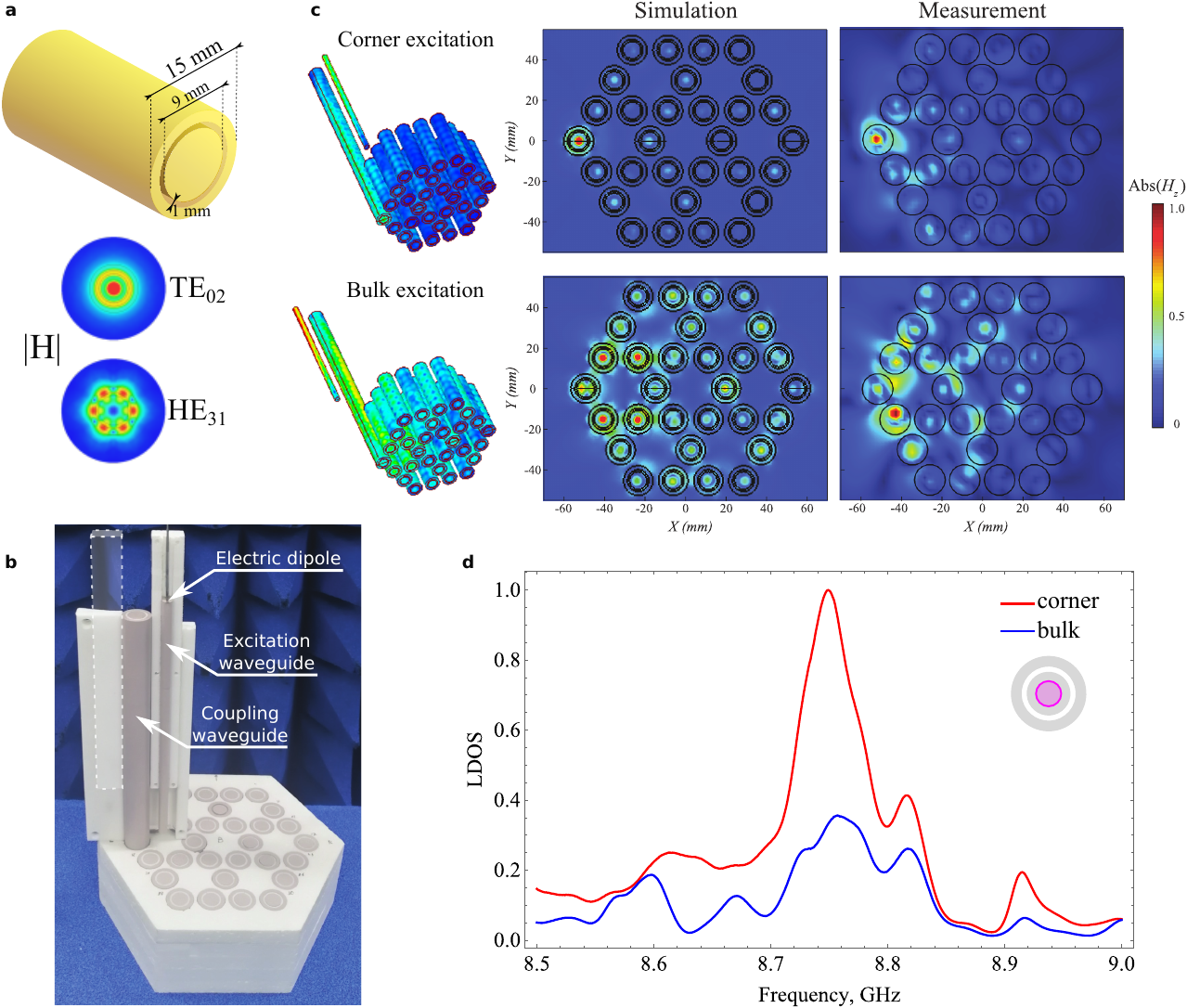}
	\caption{
		\textbf{Experimental design, field profiles, and the local density of states.} \textbf{a}, Design of a single waveguide (left) supporting TE$_{02}$ ($s$-like) and HE$_{31}$ ($f$-like) modes, with their magnetic field profiles sketched below. 
		A $1$~mm-thick cylindrical cutout is introduced in order to achieve the degeneracy of the two modes. 
		\textbf{b}, The assembled setup for the experimental near-field measurement. The array is excited by electric dipole antenna coupled to the auxiliary excitation waveguide, which has two possible positions relative to the structure (the left position is schematically shown as a white dashed area). Auxiliary waveguide is in turn coupled to the extended waveguide. The magnetic field profile is scanned by the magnetic probe located near the surface of the structure (see Supplemental Materials for details). 
		\textbf{c}, Simulated and experimental profiles of the longitudinal component of magnetic field $H_z$ for the corner and bulk states achieved for the different positions of the single-mode auxiliary excitation waveguide shown on the left. 
		The auxiliary waveguide is placed near the long coupling waveguide either closer to or further from the array center, exciting corner or bulk states of the array, respectively. 
		The frequency of the corner state is $8.740$~GHz, while the frequency of the bulk state is $8.725$~GHz. \textbf{e}, Local density of photonic states (LDOS) normalized to the maximal value retrieved from the experimental data on $H_z$ for the corner state excitation (red) and bulk state excitation (blue). The integration region is highlighted by purple on the schematics of the corner waveguide in the inset. 
	}
	\label{FigExp}
\end{figure*}

\textbf{Design of waveguide array.} To study the proposed Floquet topological insulator experimentally, we design and fabricate an array of multimode cylindrical waveguides operating in the microwave spectral range. The spectral overlapping of TE$_{02}$ and HE$_{31}$ modes with azimuthal numbers 0 and 3, respectively, is achieved by introducing cylinder-shaped air gap in an initially homogeneous cylindrical ceramic waveguide with dielectric permittivity $\eps=22$ and diameter $d_{{\rm out}} = 15$~mm as schematically shown in Fig.~\ref{FigExp}(a). Choosing the central diameter of the gap $d_{{\rm gap}}=10$~mm close to the zero of TE$_{02}$ mode electric field in the radial direction allows us to selectively perturb the dispersion of HE$_{31}$ mode keeping TE$_{02}$ mode almost unaffected. With the increase of the gap width up to $t=1$~mm HE$_{31}$ mode experiences a blue shift becoming degenerate with TE$_{02}$ mode in the frequency range around $8.7$~GHz (see details in the Supplementary Materials). The designed waveguides are further arranged in the kagome lattice with the period $a=17$~mm, which ensures the sufficient width of the band gap and efficient diffraction of the bulk modes for the chosen length of the waveguides.

Due to the nearly perfect degeneracy of the two waveguide modes, the system operates in the regime close to the limit $\kappa=\gamma=\Delta$ and therefore the localization of the corner mode is very pronounced. This allows us to efficiently excite it via the extended corner waveguide by launching a proper superposition of the waveguide modes which is achieved by exciting the extended waveguide with a properly positioned single-mode waveguide (Fig.~\ref{FigExp}c). Depending on its position, we switch between $s+f$ and $s-f$ mode superpositions required to excite the corner state or couple to the bulk modes of the array (see details in Methods and Supplementary Materials).

This possibility to tailor the phase between $s$ and $f$ orbitals enables coherent control of higher-order topological modes in our system. Generally, coherent control strategy aims to manipulate the relative phases of multiple input signals to control the system in the real time~\cite{Warren1993,Goswami2003,Rabitz2009}. In our case, this functionality is ensured by the multimode nature of the lattice and asymmetric field pattern of $s\pm f$ hybrid corner modes. The transverse electric field $\textbf{E}_\perp$ of the corner mode has maximum amplitude in the direction towards the array center featuring minimum in the opposite direction.

\textbf{Experimental results.} 
To confirm the formation of topological corner states mediated by the hybridization of TE$_{02}$ and HE$_{31}$ waveguide modes, we excite one of the doubly degenerate HE$_{11}$ modes of the auxiliary single-mode waveguide by the vertically polarized electric dipole antenna (Fig.~\ref{FigExp}b). Depending on the configuration of the setup (Fig.~\ref{FigExp}c), this launches either $s+f$ or $s-f$ superposition in the extended corner waveguide. As soon as the wave reaches the array of the waveguides, the energy  starts to redistribute across the entire array due to the coupling between the waveguides. The resulting distribution of $H_z$ component of magnetic field at the opposite side of the array is measured by a magnetic dipole antenna.

Examining the experimental results (Fig.~\ref{FigExp}c), we observe tight localization of the field when the auxiliary waveguide is brought closer to the array center and strong delocalization in the opposite case. This suggests that the corner state is indeed efficiently excited by the $s+f$ superposition in the extended waveguide, while orthogonal superposition $s-f$ excites predominantly bulk modes. In this way, we achieve coherent control of topological corner mode experimentally.

Next, we analyze the response of the array by scanning the frequency from 8.5 to 9.0~GHz with the step $1.0$~MHz and retrieving the intensity of the magnetic near field in the vicinity of the corner waveguide (see Fig.~\ref{FigExp}d and an inset). Depending on the type of excitation, we observe either a pronounced peak in the local density of states close to the frequency $8.74$~GHz or relatively monotonous behavior. The frequency of the peak corresponds to the perfect degeneracy of $s$ and $f$ modes of the isolated multimode waveguide, while the finite width of the peak suggests that the corner state persists even for imperfect mode degeneracy. This is consistent with our theoretical analysis Eq.~\eqref{eq:DetuningLimit}, which yields a range of detunings favouring the emergence of topological properties.

\section{Discussion}


In conclusion, our theoretical and experimental investigations suggest that photonic lattices with multimode meta-atoms enable novel type of multipole topology mediated by the interference of quasi-degenerate modes. Such accidental degeneracies are readily engineered by the geometric optimization of the waveguides, whereas the performance of the structure remains tolerant to the moderate detunings in the propagation constants of the modes [Eq.~\eqref{eq:DetuningLimit}].

At the same time, interaction of the modes with different symmetry is responsible for the emergence of topological bandgap. Since the coupling between the waveguide modes of different symmetry can be flexibly tailored and the degenerate modes of waveguides can emulate spin degrees of freedom, this enables extra versatility in the design of topological phenomena and generalizes the conventional spin-orbit coupling available in condensed matter systems.
%
In particular, synthetic spin-orbit coupling arising in our system involves spin-flip processes. In contrast to the conventional condensed matter systems where spin flips destroy the topological properties, here they give rise to hybrid disorder-robust topological modes. Exploiting this mechanism, we demonstrate a novel type of octupolar higher-order topology in a two-dimensional system.

From the practical perspective, our strategy enables an extreme localization of topological corner states and facilitates the formation of flat bands, the latter could be exploited to boost nonlinear phenomena. Furthermore, the multimode nature of the lattice enables coherent control of higher-order topological modes with further promising applications in topological photonics.






\section{Acknowledgments}

We acknowledge Prof.~Andrea Al{\`u} and Dr.~Dmitry Zhirihin for valuable discussions. Theoretical models were supported by the Russian Science Foundation, grant No.~20-72-10065. Numerical simulations were supported by the Priority 2030 Federal Academic Leadership Program. M.M. and M.A.G. acknowledge partial support by the Foundation for the Advancement of Theoretical Physics and Mathematics ``Basis''.

\section{Author contributions}
M.M., A.S.K. and R.S.S. contributed equally to this work. 
M.A.G. and M.M. conceived the idea. M.M. derived the analytic results. %
R.S. performed numerical simulations, designed and optimized the structure. 
A.S.K. and Z.H. assembled the experimental structure and carried out the experiments.
A.S.K., Z.H. and M.M. performed the data analysis. 
M.A.G. supervised the project. 
M.M. and M.A.G. have written the manuscript with the input from all other authors.
All authors discussed the results and contributed to the final version of the manuscript.

\section{Additional information}
Supplementary information is available in the online version of the paper. Reprints and permissions information is available online at ... . Correspondence and requests for materials should be addressed to M.A.G.

\section{Competing financial interests}
The authors declare no competing financial interests.

\bibliography{refs}

\clearpage
\newpage

\section{Methods}

\subsection{Charge density distribution}
To calculate the effective charge density distribution, we simulate the eigenmodes of a finite sample depicted schematically in Fig.~\ref{Fig2}b. Focusing on the modes below the bandgap, we sum the discrete probability distributions of the respective eigenstates: 
\begin{equation}
\sigma_{m,n,i}
=
\sum_{k = 1}^{N_\text{sites}}
\left(
\left|s_{m,n,i}^k\right|^2
+
\left|f_{m,n,i}^k\right|^2
\right)
, 
\end{equation}
where $s_{m,n,i}^k$ and $f_{m,n,i}^k$ are the amplitudes of the $s$ and $f$ modes in the site marked by the unit cell indices $m,n$ and in-cell index $i$. Index $k$ enumerates the eigenstates of the finite hexagonal sample of the lattice, and $N_\text{sites}$ is the total number of sites within the sample. Here, we assume that the eigenvectors are sorted by their eigenvalues and take into account that the total number of eigenstates is $2 N_\text{sites}$. 
Fig.~\ref{Fig2}c shows the continuous version $\sigma(x,y)$ of the discrete distribution $\sigma_{m,n,i}$ obtained by the linear interpolation. 

\subsection{Details of the experimental design}
To verify our prediction of hybrid topological corner states, we design and fabricate the experimental sample optimized for the microwave spectral range. 
The waveguides are manufactured from the commercially available low-loss high-index Taizhou TP-4 ceramics with relative permittivity $\varepsilon = 22$ and magnitude of loss $\tan\delta \simeq 10^{-3}$. These material parameters of ceramics correspond to the frequency $10$~GHz [“Taizhou Wangling”, available online at: www.wang-ling.com.cn]. 
The diameter of the auxiliary excitation waveguide is $5.48$~mm, while the two-mode waveguides consist of the inner waveguide with diameter $9$~mm and the hollow cylinder waveguide with inner and outer diameters of $11$~mm and $15$~mm, respectively. 
The height of all waveguides is equal to $50$~mm. 
An extended two-mode waveguide is assembled from the four prefabricated $50$~mm-long two-mode waveguides stacked on top of each other. 
To fill the space between the inner core and outer shell waveguides we have used the  material Rohacell HF-71 [“Rohacell”, available at: www.rohacell.com] with air-like electromagnetic properties. 
To arrange $30$ waveguides into kagome array with a period between the waveguide centers $17$~mm, the milled air-like substrate is used.

The local source of excitation~-- a subwavelength electric dipole antenna~-- is located next to the end of the auxiliary single-mode waveguide. The signal is generated by the Rohde \& Schwarz Vector Network Analyzer (VNA) ZVA50 which is connected to the antenna via $50$ Ohm coaxial cable. 
The diameter of the auxiliary waveguide is selected to support HE$_{11}$ mode at the operational frequency around $8.75$~GHz, which can be excited due to the coupling of the auxiliary waveguide to the electric dipole. 
The length of the auxiliary waveguide is $150$~mm. 
The distance between the extended and auxiliary waveguide is equal to $5$~mm, while the strength of the coupling is adjusted by moving the waveguide along its axis and changing the interaction length. 
The distance between the end of the auxiliary waveguide and the rear face of the kagome array along the $z$ axis  is $30$ mm (the spacing length SP, see Fig.~S15 in Supplementary Materials).

The near field of the array is collected with the help of a magnetic dipole antenna. The axis of the array is parallel to the $z$ axis. The dipole is placed $1$ mm above the surface of the structure and oriented along the $z$ axis. The probe moves in the $Oxy$ plane over the scanning area $140$~mm$\times 110$~mm with a $2$~mm step. 
At each sampled position, the amplitude distributions of the $z$ component of the magnetic field were collected in the frequency range of $9–10$~GHz with the step $1.0$~MHz. To avoid parasitic reflections, the measurements were performed in an anechoic chamber.

\subsection{Data availability} 
Data that are not already included in the paper and/or in the Supplementary Information are available on request from the authors.


\end{document}


\title{Supplemental Materials: \\ Multipole higher-order topology in a multimode lattice
}

\author{Maxim Mazanov}
\thanks{These three authors contributed equally}
\affiliation{School of Physics and Engineering, ITMO University, Saint Petersburg 197101, Russia}

\author{Anton S. Kupriianov}
\thanks{These three authors contributed equally}
\affiliation{State Key Laboratory of Integrated Optoelectronics, College of Electronic Science and Engineering, International Center of Future Science, Jilin University, 2699 Qianjin Street, Changchun 130012, China} 

\author{Roman S. Savelev}
\thanks{These three authors contributed equally}
\affiliation{School of Physics and Engineering, ITMO University, Saint Petersburg 197101, Russia}

\author{Zuxian~He}
\affiliation{State Key Laboratory of Integrated Optoelectronics, College of Electronic Science and Engineering, International Center of Future Science, Jilin University, 2699 Qianjin Street, Changchun 130012, China}

\author{Maxim A. Gorlach}
\email{m.gorlach@metalab.ifmo.ru}
\affiliation{School of Physics and Engineering, ITMO University, Saint Petersburg 197101, Russia}

\maketitle

\onecolumngrid

\setcounter{equation}{0}
\setcounter{figure}{0}
\setcounter{table}{0}
\setcounter{page}{1}
\setcounter{section}{0}
\makeatletter
\renewcommand{\theequation}{S\arabic{equation}}
\renewcommand{\thefigure}{S\arabic{figure}}
\renewcommand{\bibnumfmt}[1]{[S#1]}
\renewcommand{\citenumfont}[1]{S#1}

\tableofcontents

\section{Proposed geometry of the modes}

\begin{figure}[h!]
	\centering
	\includegraphics[width=0.5\textwidth]{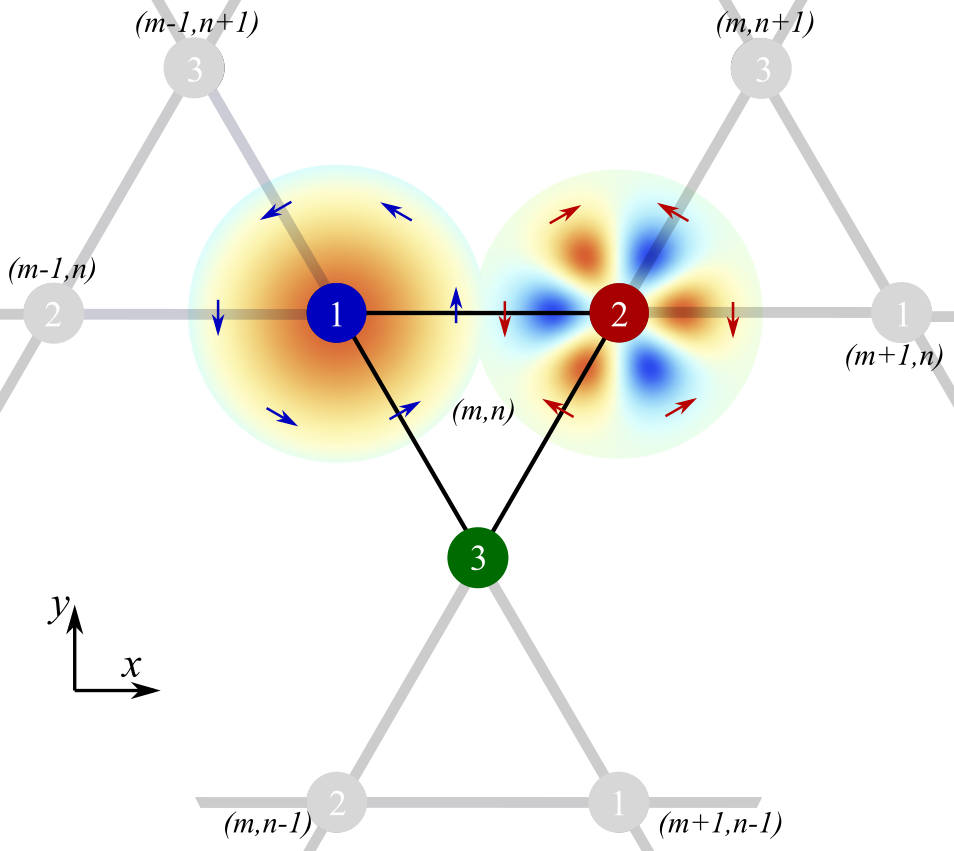}
	\caption{
		Proposed geometry of the modes: a monopolar (TE$_{0n}$) mode shown at site $ 1 $ and an octupolar (HE$_{3m}$) mode shown at site $ 2 $; arrows indicate the distribution of electric field in the respective modes. The unit cells are labeled by the indices $ (m,n) $ and the sites inside the cell are enumerated by $ \alpha\in\{1,2,3\} $. 
	}
	\label{kagome_proposed_couplings}
\end{figure}

We consider a tight-binding model for the two-mode waveguides on a kagome lattice with two degenerate (having same propagation constants at fixed frequency) modes \textit{at each site}, namely, monopolar ($s$-like) TE$_{0n}$ and octupolar ($f$-like) HE$_{3m}$ orbital modes with arbitrary radial indices $m,n$, see Fig.\ref{kagome_proposed_couplings}. 
The inter-orbital coupling is realized in this case via the in-plane component of the electric field, $\textbf{E}_\perp$. 
We note that although the $f$-mode is doubly degenerate with its inversion partner, the latter does not contribute to the effective Hamiltonian since its electric field has zero overlap integrals with the monopolar mode as dictated by the structure of the lattice. 
The qualitative picture remains the same for the choice of TM$_{0n}$ and HE$_{3m}$ orbital modes, with inter-orbital coupling realized via the out-of-plane component of the electric field, $\textbf{E}_\parallel$. 

\section{Tight-binding Bloch Hamiltonian} 

The tight-binding Bloch Hamiltonian of the proposed two-mode $s$-$f$ hybridized kagome lattice of waveguides follows from the coupled-mode theory:
\begin{eqnarray}
&&-i \frac{d}{dz} \begin{pmatrix} s \\ f \end{pmatrix}_{m,n,1}
=
\hat{\kappa}_- \begin{pmatrix} s \\ f \end{pmatrix}_{m,n,2}
+ \hat{\kappa}_+ \begin{pmatrix} s \\ f \end{pmatrix}_{m,n,3}
+ \hat{\kappa}_- \begin{pmatrix} s \\ f \end{pmatrix}_{m-1,n+1,3}
+ \hat{\kappa}_+ \begin{pmatrix} s \\ f \end{pmatrix}_{m-1,n,2}
\nonumber\\
&&-i \frac{d}{dz} \begin{pmatrix} s \\ f \end{pmatrix}_{m,n,2}
=
\hat{\kappa}_- \begin{pmatrix} s \\ f \end{pmatrix}_{m,n,3}
+ \hat{\kappa}_+ \begin{pmatrix} s \\ f \end{pmatrix}_{m,n,1}
+ \hat{\kappa}_- \begin{pmatrix} s \\ f \end{pmatrix}_{m+1,n,1}
+ \hat{\kappa}_+ \begin{pmatrix} s \\ f \end{pmatrix}_{m,n+1,3}
\nonumber\\
&&-i \frac{d}{dz} \begin{pmatrix} s \\ f \end{pmatrix}_{m,n,3}
=
\hat{\kappa}_- \begin{pmatrix} s \\ f \end{pmatrix}_{m,n,1}
+ \hat{\kappa}_+ \begin{pmatrix} s \\ f \end{pmatrix}_{m,n,2}
+ \hat{\kappa}_- \begin{pmatrix} s \\ f \end{pmatrix}_{m,n-1,2}
+ \hat{\kappa}_+ \begin{pmatrix} s \\ f \end{pmatrix}_{m+1,n-1,1}
\nonumber
,
\end{eqnarray}
where $ s $ and $ f $ denote the amplitudes of the $s$- and $f$-modes in a waveguide, respectively, and the coupling matrices read $ \hat{\kappa}_\pm = \left(
\begin{array}{cccccc}
-\kappa & \pm \Delta \\
\mp \Delta & \gamma \\
\end{array}
\right) $, where $ \kappa $, $ \gamma $, $ \Delta $ are real-valued parameters parametrizing the coupling integrals for in-plane electric fields $\textbf{E}_\perp$ of the modes $ s\Longleftrightarrow s $ ($- \kappa $), $ s\Longleftrightarrow f $ ($ \pm \Delta $), and $ f\Longleftrightarrow f $ ($ \gamma $). 
The signs of the mentioned couplings could be deduced from the symmetry of azimuthal electric field of $s$- and $f$-modes responsible for the coupling, see Fig.~\ref{kagome_proposed_couplings}. While for any coupling direction the neighbouring $s$-modes have opposite electric field directions in the coupling region and thus negative coupling (overlap integral), neighbouring $f$-modes have parallel electric field directions in the coupling region and thus positive coupling; sign of the coupling between neighbouring $s$- and $f$-modes depends on the direction of the coupling and could be deduced analogously.

\begin{figure}[h!]
	\centering
	\includegraphics[width=0.75\textwidth]{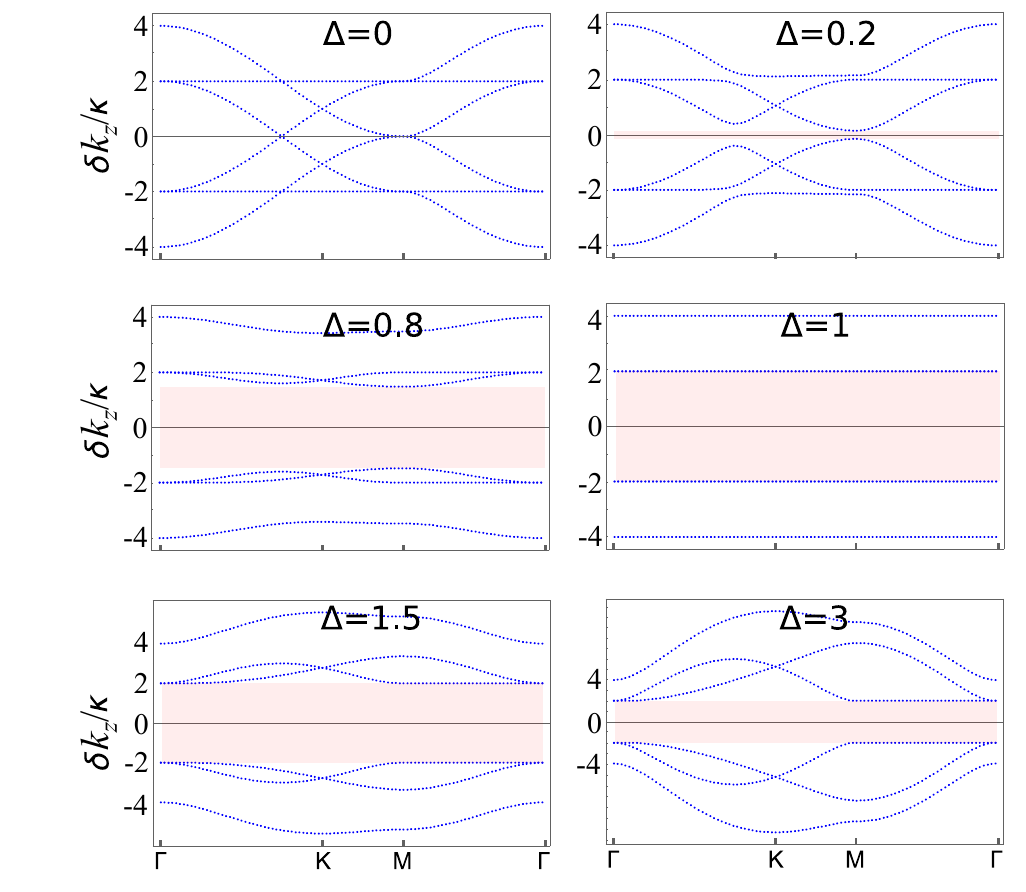}
	\caption{
		Band structure of the Hamiltonian~\eqref{Hamk} for the case $\kappa=\gamma=1$ and various $\Delta$ (shown on the panels). The bandgap is highlighted in red. 
	}
	\label{bands_transformations}
\end{figure}

The primitive lattice vectors read $ \mathbf{a} = (1,0)^T, \, \mathbf{b} = (1/2,\sqrt{3}/2)^T $, while Bloch theorem yields 
$ (s, f)^T_{m,n,\alpha} = (s, f)^T_{\alpha} e^{i \textbf{k} (m \mathbf{a} + n \mathbf{b})} \cdot e^{i \delta k_z z} $, where $ \textbf{k} = (k_x,k_y) $, $ (s, f)^T_{\alpha} $ does not depend on the cell indices, and $ \delta k_z = k_z - k_{z0} $ is the difference between propagation constants of a collective mode and of any of the two degenerate modes in an isolated waveguide. 
The Bloch Hamiltonian in this notation then reads:
\begingroup
\renewcommand*{\arraystretch}{1.0}
\setlength\arraycolsep{7pt}
\begin{equation}
\label{Hamk}
\hat{H}(\textbf{k})
=
\begin{pmatrix}
	0 & \hat{\kappa}_- + e^{-i k_x} \hat{\kappa}_+ & \hat{\kappa}_+ + e^{i \left(-\frac{kx}{2}+\frac{\sqrt{3} k_y}{2}\right)} \hat{\kappa}_- \\
	\hat{\kappa}_+ + e^{i k_x} \hat{\kappa}_- & 0 & \hat{\kappa}_- + e^{i \left(\frac{k_x}{2} + \frac{\sqrt{3} k_y}{2}\right)} \hat{\kappa}_+ \\
	\hat{\kappa}_- + e^{i \left(\frac{k_x}{2}-\frac{\sqrt{3} k_y}{2}\right)} \hat{\kappa}_+ & \hat{\kappa}_+ + e^{i \left(-\frac{k_x}{2}-\frac{\sqrt{3} k_y}{2}\right)} \hat{\kappa}_- & 0 \\
\end{pmatrix}
,
\end{equation}
\endgroup
while the eigenvalue equation reads
\begin{equation}
\label{bloch_eqn}
\hat{H}(\textbf{k})
(s_1,f_1, s_2,f_2, s_3,f_3)^T
=
\delta k_z (s_1,f_1, s_2,f_2, s_3,f_3)^T
.
\end{equation}
The main features of the band spectrum of this Hamiltonian are shown in Fig.~\ref{bands_transformations}. The bandgap opens at any nonzero $\Delta$ (which is certainly the case for any implementation of the model) at point $\tilde{K}$ between on the $K$-$M$ line (e.g., $\tilde{K} = \{ \pi/2 , \sqrt{3}\pi/2 \}$)
and $M$ point. 
More precisely, for $\Delta=0$ the $\tilde{K}$ point lies in the middle of the whole continuous line of degeneracy (rather than being a Dirac point).
At $\kappa=\gamma=\Delta$, the band structure consists of \textit{six fully flat bands}, thus the corresponding eigenstates are fully localized. For larger $\Delta$, the width of the bandgap remains the same as in the flat-band case.

\section{Generalized chiral symmetry}
The Bloch Hamiltonian \eqref{Hamk} written for the three-cite unit cell possesses generalized chiral symmetry for any choice of couplings. The proof is given simply by constructing a respective unitary generalized chiral symmetry operator, 
\begin{eqnarray}
\hat{\Gamma}_3
=
\begin{pmatrix}
	\hat{I} & 0 & 0 \\
	0 & e^{i 2\pi/3} \cdot \hat{I} & 0 \\
	0 & 0 & e^{i 4\pi/3} \cdot \hat{I} \\
\end{pmatrix}
,
\end{eqnarray}
where $\hat{I}$ is a $2\times 2$ unit matrix, and zeros stand for the $2\times 2$ zero matrices. Then, it is straightforward to check that 
\begin{eqnarray}
\hat{H}(\textbf{k})
+
\hat{\Gamma}_3
\hat{H}(\textbf{k})
\hat{\Gamma}_3^{-1} 
+
\hat{\Gamma}_3 \hat{\Gamma}_3 
\hat{H}(\textbf{k})
\hat{\Gamma}_3^{-1} \hat{\Gamma}_3^{-1} 
=
0
,
\end{eqnarray}
which proves that $\hat{H}(\textbf{k})$ indeed possesses generalized chiral symmetry. This explains why the spectrum is symmetric, and implies that all corner states may be obtained from a single one by successively applying the generalized chiral symmetry operator $\hat{\Gamma}_3$ to one known eigenstate.

\section{Equivalent two-layer one-mode model}
\begin{figure}[h!]
	\centering
    \includegraphics[width=0.60\textwidth]{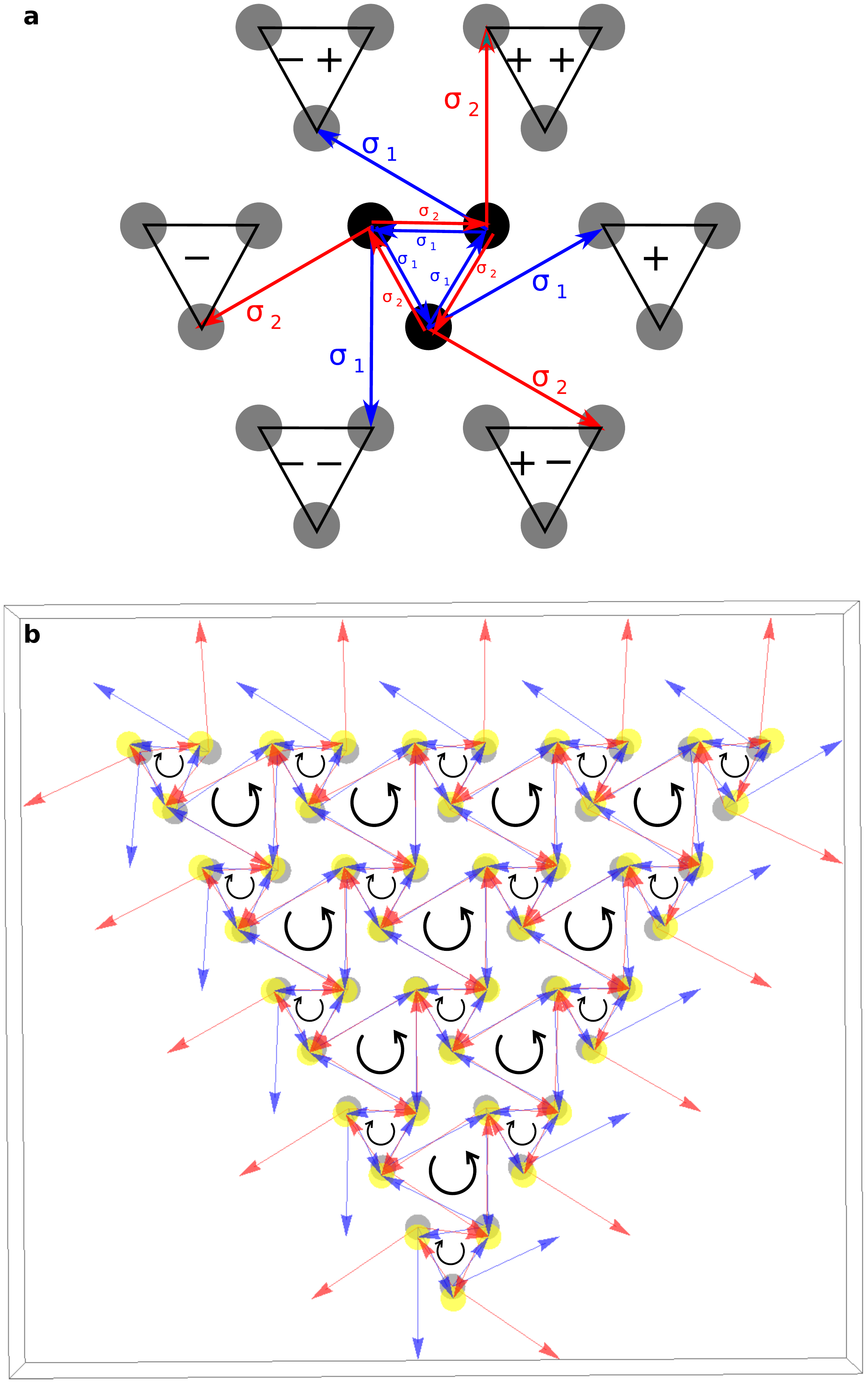} 
    \caption{Equivalent two-layer model. (a) Possible geometry of couplings deduced from the block in Eq.~\eqref{hsmall}, (b) Top view of a two-layer model. }%
	\label{eqauivalent_bands}
\end{figure}

At $ \kappa = \gamma $, with the help of unitary transformation
\begin{eqnarray}
\label{Ubasis_two_layer}
\hat{U}
=
\frac{1}{\sqrt{2}}\left(\begin{array}{cccccc}
 0 & 0 & -i & 0 & 0 & 1 \\
 0 & 0 & -i & 0 & 0 & -1 \\
 -i & 0 & 0 & 1 & 0 & 0 \\
 -i & 0 & 0 & -1 & 0 & 0 \\
 0 & -i & 0 & 0 & 1 & 0 \\
 0 & -i & 0 & 0 & -1 & 0 \\
\end{array}\right)
\end{eqnarray}
the Bloch Hamiltonian~\eqref{Hamk} takes the form
\begin{eqnarray}
\label{H1}
\hat{H}_1
=
\hat{U}^{\dag} \hat{H} \hat{U}
=
\left(\begin{array}{cc}
0 & \hat{h}(\boldsymbol{k}) \\
\hat{h}^{\dagger}(\boldsymbol{k}) & 0
\end{array}\right)
, 
\end{eqnarray}
where each of the two off-diagonal blocks read 
\begin{eqnarray}
\label{hsmall}
- i \hat{h}^{\dag}(\boldsymbol{k}) 
=
\left(\begin{array}{ccc}
0 & \sigma_1 (--) + \sigma_2 & \sigma_1 + \sigma_2 (-) \\
\sigma_1 + \sigma_2 (++) & 0 & \sigma_1 (-+) + \sigma_2 \\
\sigma_1 (+) + \sigma_2 & \sigma_1 + \sigma_2 (+-) & 0
\end{array}\right)
. 
\end{eqnarray}
Here, we have shortened the Bloch factors for convenience: 
\begin{eqnarray}
&&(\pm_{(1)},\pm_{(2)}) \equiv \exp\left[{i\left(\pm_{(1)} k_x /2 \pm_{(2)} \sqrt{3} k_y / 2\right)}\right]
,\\
&&(\pm) \equiv \exp\left[{\pm i k_x}\right]
, 
\end{eqnarray}
and introduced the parameters
\begin{eqnarray}
&&\sigma_1
=
\kappa - \Delta
,\\
&&\sigma_2
=
\kappa + \Delta
. 
\end{eqnarray}
The form of the Hamiltonian \eqref{H1} suggests that our two-mode single-layer model is equivalent to some \textit{two-layer single-mode model} with layers stacked one upon another, with \textit{(purely imaginary) couplings of the sites arising only between the layers}. Further, the block~\eqref{hsmall} hints at the possible exact realization of such a model, see Fig.~\ref{eqauivalent_bands} (showing the top view of the structure). In this figure, the lower kagome layer is coloured gray, the upper~-- yellow, and the arrows indicate the couplings between the layers: blue arrows represent $\sigma_1$, and red arrows represent $\sigma_2$ shown only for the $\hat{h}^{\dagger}(\boldsymbol{k})$ block. Due to the Hermitian nature of the model, the coupling links are reciprocal. 

Importantly, the model features effective ``chirality'' (opposite for two types of hoppings~-- between layers $1\rightarrow 2$ and $2\rightarrow 1$) when $\sigma_1 \neq \sigma_2$, i.e. when the inter-mode coupling $\Delta \neq 0$.
The ``chirality'' is schematically represented in Fig.~\ref{eqauivalent_bands} by the curved arrows for the case $ \Delta > 0 $. 
Such ``chirality'' could be also interpreted as a ``flux per plaquette'', however, such that the net flux per unit cell vanishes, similar to e.g. Haldane model~\cite{Haldane}. 
In fact, if we associate the two layers with two pseudospins, this effective chirality induced by nonzero $\Delta$ could be interpreted as a \textit{spin-orbit coupling which involves spin flip} (see also the next section).

Finally, in the general case $ \kappa \neq \gamma $, the Bloch Hamiltonian~\eqref{Hamk} after the same transform takes the similar form, but with additional diagonal terms: 
\begin{eqnarray}
\label{H1}
\hat{H}_1
=
\left(\begin{array}{cc}
\hat{h}_\text{intra}(\boldsymbol{k}) & \hat{h}(\boldsymbol{k}) \\
\hat{h}^{\dagger}(\boldsymbol{k}) & \hat{h}_\text{intra}(\boldsymbol{k})
\end{array}\right)
, 
\end{eqnarray}
where the intra-layer couplings read 
\begin{eqnarray}
\label{hintra}
\hat{h}_\text{intra}(\boldsymbol{k})
=
\frac{\kappa -\gamma}{2} \left(
\begin{array}{ccc}
 0 & 1+e^{\frac{1}{2} i \left(k_x+\sqrt{3} k_y\right)} & 1+e^{i k_x} \\
 1+e^{-\frac{1}{2} i \left(k_x+\sqrt{3} k_y\right)} & 0 & 1+e^{\frac{1}{2} i \left(k_x-\sqrt{3} k_y\right)} \\
 1+e^{-i k_x} & 1+e^{\frac{1}{2} i \left(\sqrt{3} k_y-k_x\right)} & 0 \\
\end{array}
\right)
. 
\end{eqnarray}
These intra-layer couplings are similar to the single-layer kagome model with equal intra- and inter-cell couplings. The latter single-layer model is itself not topological, which stresses once more that in the equivalent two-layer model, topology is related exclusively to the special structure of inter-layer couplings.

\section{Effective spin-flip spin-orbit coupling}
In this section, we discuss the effective spin-flip spin-orbit coupling (SOC) which opens the topological gap in our model and compare it to the conventional next-nearest neighbour spin-preserving spin-orbit coupling terms in a single-mode kagome lattice similar to one used in Haldane model~\cite{SOC-Kagome,Haldane}. 

We first focus on the gap opening at the Dirac-like point $\tilde{K} = \{ \pi/2 , \sqrt{3}\pi/2 \}$ which at $\Delta=0$ lies in the middle of a continuous line of degeneracy. Notice that three gaps open simultaneously: at zero $\delta k_z$, between first and second dispersion branches, and also between fifth and sixth bands.

To grasp the connection of the effective spin-orbit coupling to the zero-energy topological gap, we diagonalize the Hamiltonian~\eqref{Hamk} near the $\tilde{K}$ point, and then expand it up to liner terms in $\{ \delta k_x, \delta k_y \} = \{ k_x - \pi/2, k_y - \sqrt{3}\pi/2 \}$. Then, we interpret the off-diagonal part of the difference $\hat{H}^{eff}_{SO}\equiv \hat{H}(\tilde{K} + \delta k_x, \delta k_y) - \hat{H}(\tilde{K})$ as an effective spin-orbit interaction: 
\begin{eqnarray}
\label{SOC}
\hat{H}^{eff}_{SO}
=
&&\delta k_x \left(
\begin{array}{cccccc}
 0 & -\frac{\kappa }{2} & \frac{3 \Delta }{2 \sqrt{2}} & \frac{(-3-7 i) \Delta -(12-4 i) \kappa }{16 \sqrt{2}} & -\frac{3 \Delta }{2 \sqrt{2}} & \frac{(-3-7 i) \Delta +(12-4 i) \kappa}{32} \\
   & 0 & \frac{(3+7 i) \Delta -(12-4 i) \kappa}{32}  & -\frac{3 \Delta }{2 \sqrt{2}} & \frac{(3+7 i) \Delta +(12-4 i) \kappa}{32} & \frac{3 \Delta }{2 \sqrt{2}} \\
   &    & 0 & \frac{\kappa }{4} & \frac{- \Delta + 6 i \kappa}{8} & \frac{2 \kappa -3 \Delta}{8} \\
   &    &    & 0 & \frac{3 \Delta +2 \kappa}{8} & \frac{\Delta +6 i \kappa}{8}  \\
   &    &    &    & 0 & \frac{\kappa }{4} \\
   &    &    &    &    & 0 \\
\end{array}
\right)
+   \nonumber \\
&&\sqrt{\frac{3}{2}} \delta k_y
\left(
\begin{array}{cccccc}
 0 & -\kappa  & -\frac{\Delta }{2} & \frac{(1-7 i) \Delta +(4+4 i) \kappa }{16} & \frac{\Delta }{2} & \frac{(1-7 i) \Delta -(4+4 i) \kappa}{16} \\
 & 0 & \frac{(-1+7 i) \Delta +(4+4 i) \kappa}{16} & \frac{\Delta }{2} & \frac{(-1+7 i) \Delta -(4+4 i) \kappa}{16} & -\frac{\Delta }{2} \\
 &   & 0 & \frac{\kappa }{2} & -\frac{\Delta +2 i \kappa}{4} & \frac{2 \kappa -3 \Delta}{4} \\
 &   &   & 0 & \frac{3 \Delta +2 \kappa}{4} & \frac{\Delta -2 i \kappa}{4} \\
 &   &   &   & 0 & \frac{\kappa }{2} \\
 &   &   &   &   & 0 \\
\end{array}
\right)
\end{eqnarray}
(here, the lower triangle of both matrices is understood to be filled such that the matrices are Hermitian). Since the result could not be expanded in tensor products of spin-1 (Pauli) and spin-2 matrices, we instead use the expansion into tensor products of Pauli ($\sigma_i$) and Gell-Mann matrices $\lambda_i$, the latter parameterize all Hermitian $3\times3$ matrices: 
\begin{eqnarray}
&&\lambda_0 = 
\frac{1}{\sqrt{3}}
\left(
\begin{array}{ccc}
 1 & 0 & 0 \\
 0 & 1 & 0 \\
 0 & 0 & 1 \\
\end{array}
\right)
,
\lambda_1 = 
\frac{1}{\sqrt{2}}
\left(
\begin{array}{ccc}
 0 & 1 & 0 \\
 1 & 0 & 0 \\
 0 & 0 & 0 \\
\end{array}
\right)
,
\lambda_2 = 
\frac{1}{\sqrt{2}}
\left(
\begin{array}{ccc}
 0 & -i & 0 \\
 i & 0 & 0 \\
 0 & 0 & 0 \\
\end{array}
\right)
,
\lambda_3 = 
\frac{1}{\sqrt{2}}
\left(
\begin{array}{ccc}
 1 & 0 & 0 \\
 0 & -1 & 0 \\
 0 & 0 & 0 \\
\end{array}
\right)
,
\lambda_4 = 
\frac{1}{\sqrt{2}}
\left(
\begin{array}{ccc}
 0 & 0 & 1 \\
 0 & 0 & 0 \\
 1 & 0 & 0 \\
\end{array}
\right)
, \nonumber \\
&&\lambda_5 = 
\left(
\begin{array}{ccc}
 0 & 0 & -i \\
 0 & 0 & 0 \\
 i & 0 & 0 \\
\end{array}
\right)
,
\lambda_6 = 
\frac{1}{\sqrt{2}}
\left(
\begin{array}{ccc}
 0 & 0 & 0 \\
 0 & 0 & 1 \\
 0 & 1 & 0 \\
\end{array}
\right)
,
\lambda_7 = 
\frac{1}{\sqrt{2}}
\left(
\begin{array}{ccc}
 0 & 0 & 0 \\
 0 & 0 & -i \\
 0 & i & 0 \\
\end{array}
\right)
,
\lambda_8 = 
\frac{1}{\sqrt{6}}
\left(
\begin{array}{ccc}
 1 & 0 & 0 \\
 0 & 1 & 0 \\
 0 & 0 & -2 \\
\end{array}
\right)
. \nonumber
\end{eqnarray}
Thus, the effective SOC $\hat{H}^{eff}_{SO}$ reads 
\begin{eqnarray}
\hat{H}^{eff}_{SO}
=
\sum_{i = 0}^{8}
\sum_{j = 0}^{3}
C_{ij}
\hat{\sigma}_j \otimes \hat{\lambda}_i
. 
\end{eqnarray}
Although the matrix of coefficients $C_{ij}$ obtained from the series expansion has a complicated form, most importantly, it features nonzero elements besides $C_{1i}$ and $C_{4i}$ which correspond to contributions to the Hamiltonian proportional to $\hat{\sigma}_0$ or $\hat{\sigma}_3$. This also holds for any permutation of the basis of the diagonalized SOC Hamiltonian~\eqref{SOC}. 
Thus, the effective spin-orbit coupling in our model includes spin-flip couplings.

For the sake of comparison, we recall the form of the conventional next-nearest neighbour spin-orbit coupling terms which are typically introduced into single-mode spinful kagome lattices~\cite{SOC-Kagome}. The structure of such spin-orbit couplings is similar to one used in Haldane model and, in sharp contrast to the effective SOC in our model, it involves two diagonal $3\times3$-blocks for each spin and thus no couplings are accompanied by the spin flip. 
This conclusion persists if, as previously, we diagonalize the Hamiltonian near $K (K')$ point which will retain the block-diagonal structure of the couplings, thus spin-flip couplings will also be absent in this basis. 

We conclude that \textit{there is a crucial distinction between the conventional SOC and the effective SOC in our model which necessarily features spin-flip coupling}. 

Interestingly, the nature of unusual spin-flip couplings in our model could be understood  more intuitively in the basis~\eqref{Ubasis_two_layer} introduced in the previous section. As we have shown, in this basis the Hamiltonian features two types of off-diagonal terms (both nearest-neighbour and next-nearest-neighbour): spin-flip coupling terms (equal to $\kappa$), and \textit{spin-flip spin-orbit coupling terms} proportional to $\Delta$.

This clarifies that the mechanism of topological gap opening in our model is connected precisely to such spin-flip spin-orbit coupling terms. Such type of couplings, to the best of our knowledge, has not been previously associated to topology in condensed-matter physics, but rather treated as an undesirable spin-mixing perturbation (as e.g. in Kane-Mele model~\cite{Kane-Mele}).

\section{Rotation operator eigenvalues at high-symmetry points}

In this section, we determine the topological class of our two-mode lattice. We adopt the classification of the higher-order topological crystalline insulators with time-reversal and $ C_n $ symmetry~\cite{HOTCIs}. In our case the lattice possesses $ C_3 $ symmetry, therefore the topological class is described by two topological indexes, $ [K^{(3)}_1] $ and $ [K^{(3)}_2] $. Here, $ [K^{(3)}_p] = \# K^{(3)}_p - \# \Gamma^{(3)}_p $, and $ \# \Pi^{(3)}_p $ is the number of energy bands below the gap at a particular high-symmetry point (HSP) $ \boldsymbol{\Pi}^{(3)} $ which have the eigenvalue $ \Pi^{(3)}_p = e^{2\pi i (p-1)/3} $ with respect to a threefold rotation operator $ \hat{r}_3 $.

We first choose the three-site elementary cell as depicted in Fig.\ref{kagome_proposed_couplings}~(a) by the blue sites. Then, the rotation operator reads 
\[
\hat{r}_3
=
\begin{pmatrix}
	0 & 0 & 1 & 0 & 0 & 0 \\
	0 & 0 & 0 & 1 & 0 & 0 \\
	0 & 0 & 0 & 0 & 1 & 0 \\
	0 & 0 & 0 & 0 & 0 & 1 \\
	1 & 0 & 0 & 0 & 0 & 0 \\
	0 & 1 & 0 & 0 & 0 & 0 \\
\end{pmatrix}
.
\]
Note that this form of the operator is consistent with observation that the chosen waveguide modes are also invariant under $C_3$ rotation. 

It is then straightforward to check that, indeed, the Bloch Hamiltonian commutes with rotation operator at HSPs, i.e., $ [\hat{r}_3,\hat{H}(\boldsymbol{\Gamma})]=0 $ and $ [\hat{r}_3,\hat{H}(\textbf{K})]=0 $, which implies that the eigenvectors of Hamiltonian at HSPs are simultaneously the eigenvectors of the rotation operator. 
Hence, at first for the simpler case $ \kappa=\gamma > 0 $, we find the eigenvectors $ |\psi_i\rangle $ of $ \hat{H}(\textbf{K}) $ and $ \hat{H}(\boldsymbol{\Gamma}) $ which transform as irreducible representations of $ C_3 $ group, and then obtain their eigenvalues with respect to rotation operator as $ \lambda_i = \langle \psi_i | \hat{r}_3 \psi_i \rangle / \langle \psi_i | \psi_i \rangle $. We obtain the following eigenvalues (states are sorted by energy):

$\bullet$ for the $ \textbf{K}_1 $ point: $ e^{\frac{2\pi}{3}i} $ (lowest band), $ \{e^{\frac{4\pi}{3}i}, \, 1\} $ (degenerate bands with negative energy), $ \{e^{\frac{4\pi}{3}i}, \, 1\} $ (degenerate bands with positive energy), $ e^{\frac{2\pi}{3}i} $ (highest band).

$\bullet$ for the $ \textbf{K}_2 $ point: $ e^{\frac{4\pi}{3}i} $ (lowest band), $ \{e^{\frac{2\pi}{3}i}, \, 1\} $ (degenerate bands with negative energy), $ \{e^{\frac{2\pi}{3}i}, \, 1\} $ (degenerate bands with positive energy), $ e^{\frac{4\pi}{3}i} $ (highest band).

$\bullet$ for the $ \boldsymbol{\Gamma} $ point: $ 1 $ (lowest band), $ \{e^{\frac{2\pi}{3}i}; \, e^{\frac{4\pi}{3}i}\} $ (degenerate bands with negative energy), $ \{e^{\frac{2\pi}{3}i}; \, e^{\frac{4\pi}{3}i}\} $ (degenerate bands with positive energy), $ 1 $ (highest band).

Therefore, we obtain the following topological invariants: $ [K^{(3)}_1] = 0 $ and $ [K^{(3)}_2] = 0 $. For crystalline insulators with sites retained at corners, this would imply vanishing bulk polarization and corner charge.

\begin{figure}[h!]
	\centering
	\includegraphics[width=0.6\textwidth]{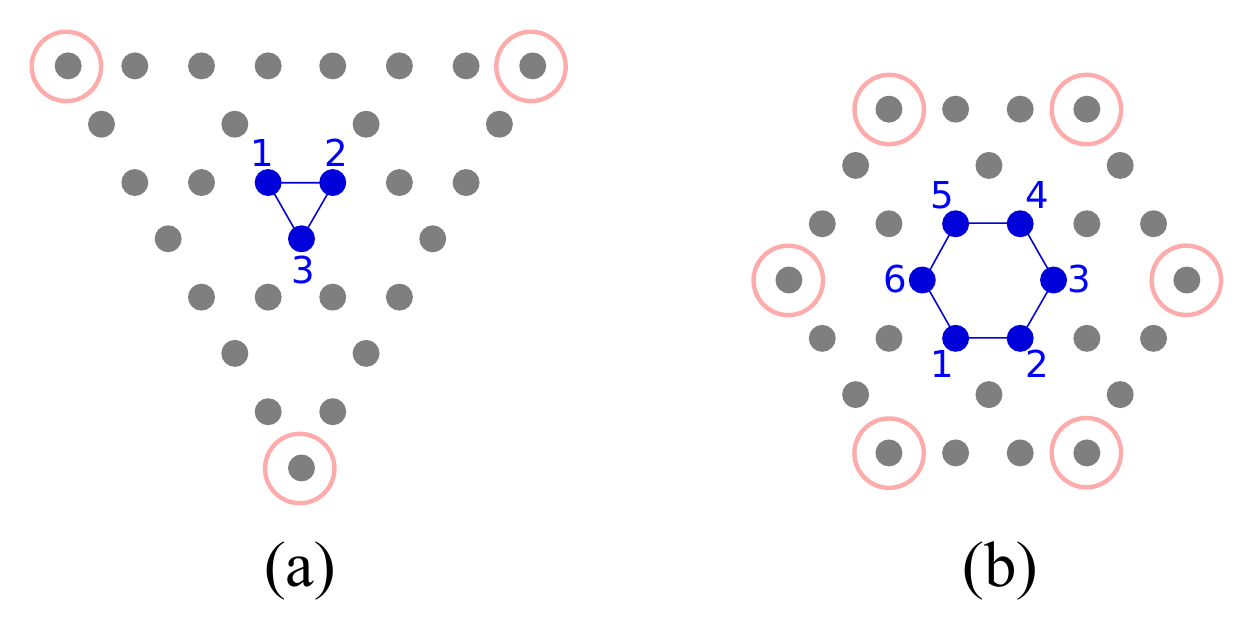}
	\caption{
		Finite kagome lattices: (a) triangular with corner sites, and (b) hexagonal without corner sites, and the corresponding unit cells. 
	}
	\label{retained_corner_sites}
\end{figure}

We then try another, six-site elementary cell as depicted in Fig.\ref{kagome_proposed_couplings}~(b). Then, the rotation operator $\hat{r}_3$ reads like $\hat{r}'_3$ if all the nulls are replaced by $2\times2$ null-matrices, and $1$s are replaced by $2\times2$ identity matrices. 
This form of the operator is also consistent with the symmetry of waveguide modes. The eigenvectors for which we will find the $\hat{r}'_3$ eigenvalues could be either found from the corresponding $12\times 12$ Hamiltonian, or constructed from the eigenvectors of $6\times 6$ Hamiltonian. We follow the latter way, and construct the eigenstates according to Bloch theorem. For any $6\times 6$ Hamiltonian $3$-site eigenvector $\mathbf{v}_3 = \{s_1, f_1, s_2, f_2, s_3, f_3\}$, the constructed $6$-site eigenvector reads
\begin{eqnarray}
\mathbf{v}_6
&=&
\{
s_1, f_1, s_2, f_2, 0, 0, 0, 0, 0, 0, 0, 0
\}
+\nonumber\\
&+& e^{i \left(\frac{k_x}{2}+\frac{\sqrt{3} k_y}{2}\right)} 
\cdot
\{
0, 0, 0, 0, s_3, f_3, s_1, f_1, 0, 0, 0, 0
\}
+\nonumber\\
&+& e^{i \left(\frac{-k_x}{2}+\frac{\sqrt{3} k_y}{2}\right)}
\cdot
\{
0, 0, 0, 0, 0, 0, 0, 0, s_2, f_2, s_3, f_3
\}
. 
\end{eqnarray}
Proceeding as in the case of three-site cell, we find that the rotation operator $\hat{r}'_3$ eigenvalues are \textit{the same} for the $ \textbf{K} $ and $ \boldsymbol{\Gamma} $ points; namely: $ 1 $ (lowest band), $ \{e^{\frac{2\pi}{3}i}; \, e^{\frac{4\pi}{3}i}\} $ (degenerate bands with negative energy), $ \{e^{\frac{2\pi}{3}i}; \, e^{\frac{4\pi}{3}i}\} $ (degenerate bands with positive energy), $ 1 $ (highest band).
Therefore, we again obtain the following topological invariants: $ [K^{(3)}_1] = 0 $ and $ [K^{(3)}_2] = 0 $, hence the associated bulk polarization vanishes. 

\begin{figure}[h!]
	\centering
	\includegraphics[width=0.65\textwidth]{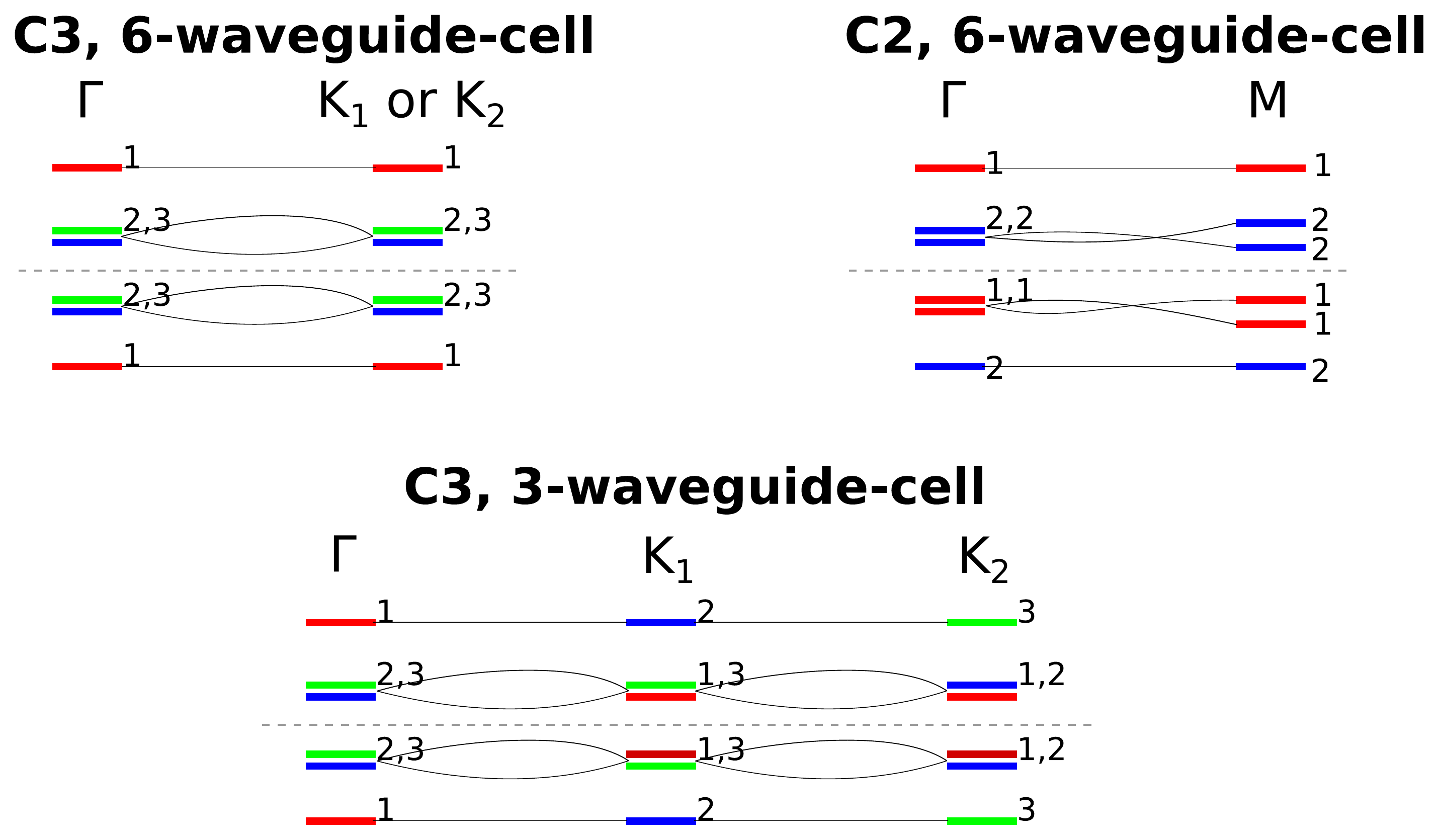}
	\caption{
	    Eigenvalues at HSP (numbers label the integer $p$ in the eigenvalues, $ \Pi^{(3)}_p = e^{2\pi i (p-1)/3} $). 
	}
	\label{eigenvalues}
\end{figure}

The eigenvalues are summarized in Fig.\ref{eigenvalues}. 

\section{Possibility of bulk quadrupole moment}
In this section, we derive the symmetry constraints on possible values of bulk quadrupole moments in a system with $C_3$ symmetry and vanishing bulk polarization. 

The quadrupole moment in 2D has four components, of which only two are independent due to the fact that the quadrupole moment is symmetric and could be brought to traceless form. Thus, the generic form of quadrupole moment in 2D is the following: 
\begin{equation}
    \hat{q}
    =
    \left(
    \begin{array}{cc}
     a & b \\
     b & -a \\
    \end{array}
    \right)
    . 
\end{equation}
The $C_3$-rotation matrix reads 
\begin{equation}
    \hat{T}_{3}
    =
    \left(
    \begin{array}{cc}
     -1/2 & -\sqrt{3}/2 \\
     \sqrt{3}/2 & -1/2 \\
    \end{array}
    \right)
    . 
\end{equation}
Thus, the rotated quadrupole moment $\hat{q}' = \hat{T}_{3}^\dag \hat{q} \hat{T}_{3}$ reads
\begin{equation}
    \hat{q}'
    =
    \frac{1}{2} 
    \left(
    \begin{array}{cc}
    -a - \sqrt{3} b & \sqrt{3} a-b \\
    \sqrt{3} a-b & a + b \sqrt{3} \\
    \end{array}
    \right)
    .
\end{equation}
The $C_3$ and the point-group symmetry of our model constrains this rotated form to be equivalent to $\hat{q}$ up to the sum of four possible Kronecker products $\mathbf{a}_i \otimes \mathbf{a}_j$ constructed from the lattice vectors $\mathbf{a}_i$ weighted by integer coefficients $n_{ij},\, i,j \in \{1,2\}$. We can choose the lattice vectors as $\mathbf{a}_1 = \{ 1, 0 \}$, $\mathbf{a}_2 = \{ 1/2, \sqrt{3}/2 \}$. In this basis, the sum of four Kronecker products reads 
\begin{equation}
    \delta \hat{q}_{ij}
    =
    \sum_{i=1}^{4} n_{ij} \mathbf{a}_i \otimes \mathbf{a}_j
    =
    \left(
    \begin{array}{cc}
    n_{11} + n_{12}/2 + n_{21}/2 + n_{22}/4 & \sqrt{3} n_{12}/2 + \sqrt{3} n_{22}/4 \\
    \sqrt{3} n_{21}/2 + \sqrt{3} n_{22}/4 & 3 n_{22}/4 \\
    \end{array}
    \right)
    .
\end{equation}
The solutions of $\hat{q} = \hat{q}' + \delta\hat{q}$ with integer $n_{ij}$ read simply $a \in \mathbb{Z},\, b = \sqrt{3} \cdot \mathbb{Z}$. 
Thus, the possible quadrupole moment in systems with $C_3$ symmetry reads 
\begin{equation}
    \hat{q}_{C_3}
    =
    \left(
    \begin{array}{cc}
     \alpha & \sqrt{3} \beta \\
     \sqrt{3} \beta & -\alpha \\
    \end{array}
    \right) 
\end{equation}
for any integers $\alpha,\beta  \in \mathbb{Z}$.

\section{Numerical calculation of multipole moments}

The universally accepted method to calculate the quadrupole and octupole moments in crystalline insulators is via the nested Wilson Loops calculation in the wavevector space~\cite{Benalcazar_2017_Science,Benalcazar_2017_PRB}. However, to date it has been applied only to $C_4$- and $C_2$-symmetric two-dimensional quadrupole insulators, while the modification of such derivation is not straightforward for $C_3$- and $C_6$-symmetric lattices as they possess non-orthogonal translation vectors. 
Despite this difficulty, one could use the definition of the ``real-space'' quadrupole moment (as well as other multipole moments) for the finite lattices with periodic or open boundary conditions. This definition is widely used in the investigation of topological Anderson and amorphous insulators~\cite{Li2020Oct,Agarwala2020Mar} where the nested Wilson loop procedure could not be applied. For the case of half-filled bands, the definition reads:
\begin{eqnarray}
\label{real-space moments}
q_{\alpha_1 \alpha_2 ...}
=
\frac{1}{2 \pi} \operatorname{Im} \log \left[\operatorname{det}\left(U^{\dagger} \hat{Q} U\right) \sqrt{\operatorname{det}\left(\hat{Q}^{\dagger}\right)}\right]
, 
\end{eqnarray}
where $q_{\alpha_1 \alpha_2 ...}$ is the desired component real-space multipole moment (e.g. the off-diagonal component of the quadrupole moment, $q_{xy}$), $\hat{Q} = \exp[2 \pi i \hat{q}_{\alpha_1 \alpha_2 ...}]$, $\hat{q}_{\alpha_1 \alpha_2 ...}$ is the corresponding multipole moment operator (e.g. $\hat{q}_{xy} = \hat{x} \hat{y} / S$, where $\hat{x}$, $\hat{y}$ are the position operators in $x$ and $y$ directions, respectively, and $S$ is the total area of the sample), and the unitary matrix $\hat{U}$ is constructed by the column-wise arrangement of the occupied eigenstates (such that $\hat{U}^\dag \hat{U} = \hat{1}$, and $\hat{U} \hat{U}^\dag$ is the projector onto the occupied eigenstates). 
The formula Eq.~\eqref{real-space moments} is constructed as a generalization of the Resta's formula of polarization~\cite{Resta1998Mar}, while $\sqrt{\operatorname{det}\left(\hat{Q}^{\dagger}\right)}$ corresponds to the subtracted value of the calculated multipole moment in the half-filled limit of evenly distributed probability (the ``atomic limit'').

The multipole moments in two dimensions are defined via the multipolar decomposition of the effective potential $\phi(r,\theta)$, where $r$ and $\theta$ are radial and polar coordinates in the 2D cylindrical coordinate system: 
\begin{eqnarray}
\phi(r,\theta)
=
a_0 \ln \frac{1}{r} 
+
\sum_{n=1}^{\infty} \frac{a_n \cos(n \theta) + b_n \sin(n \theta)}{r^n},
\end{eqnarray}
where the coefficient $a_0$ describes the net effective charge of the object, while coefficients $a_n$ and $b_n$ describe the multipoles of $n^{\text{th}}$ order, and read: 
\begin{eqnarray}
a_n = \frac{1}{n} \iint dS\, r^n \cos(n \theta) \rho(r,\theta)
, \quad
b_n = \frac{1}{n} \iint dS\, r^n \sin(n \theta) \rho(r,\theta)
,
\end{eqnarray}
where $\rho(r,\theta)$ is the effective charge distribution. For instance, in the $1^{\text{st}}$ order $a_1$ and $b_1$ are $x$ and $y$ polarization components, respectively. In the $2^{\text{nd}}$ order $a_2$ and $b_2$ are the quadrupole moment components, which could be rewritten as 
\begin{eqnarray}
a_2 = \frac{1}{2} \iint dx dy\, (x^2 - y^2) \rho(x,y)
, \quad
b_2 = \iint dx dy\, xy \rho(x,y)
. 
\end{eqnarray}
Thus, to calculate the respective components of the quadrupole moment via Eq.~\eqref{real-space moments}, we need to take operators $\hat{q}_1 \equiv (\hat{x}^2 - \hat{y}^2)/(2 S) $ and $\hat{q}_2 \equiv 2 \hat{x} \hat{y} /S$ as $\hat{q}_{\alpha_1 \alpha_2 ...}$. 
Note that for the canonical quadrupole insulator~\cite{Benalcazar_2017_PRB}, \textit{both} $q_1$ and $q_2$ are quantized to value $0.5$.  

By analogy, the third-order, or ``2D-octupolar'', moment components could be calculated by rewriting $a_3$ and $b_3$ as 
\begin{eqnarray}
a_3 = \frac{1}{3} \iint dx dy\, (x^3 - 3 x y^2) \rho(x,y)
, \quad
b_3 = \frac{1}{3} \iint dx dy\, (3 x^2 y - y^3) \rho(x,y)
, 
\end{eqnarray}
thus we need to take the operators $\hat{o}_1 \equiv (\hat{x}^3 - 3 \hat{x} \hat{y}^2)/(3 S^{3/2}) $ and $\hat{o}_2 \equiv (3 \hat{x}^2 \hat{y} - \hat{y}^3)/(3 S^{3/2})$ as $\hat{q}_{\alpha_1 \alpha_2 ...}$ in Eq.~\eqref{real-space moments} in this case.

\begin{figure}[h!]
	\centering
	\includegraphics[width=0.99\textwidth]{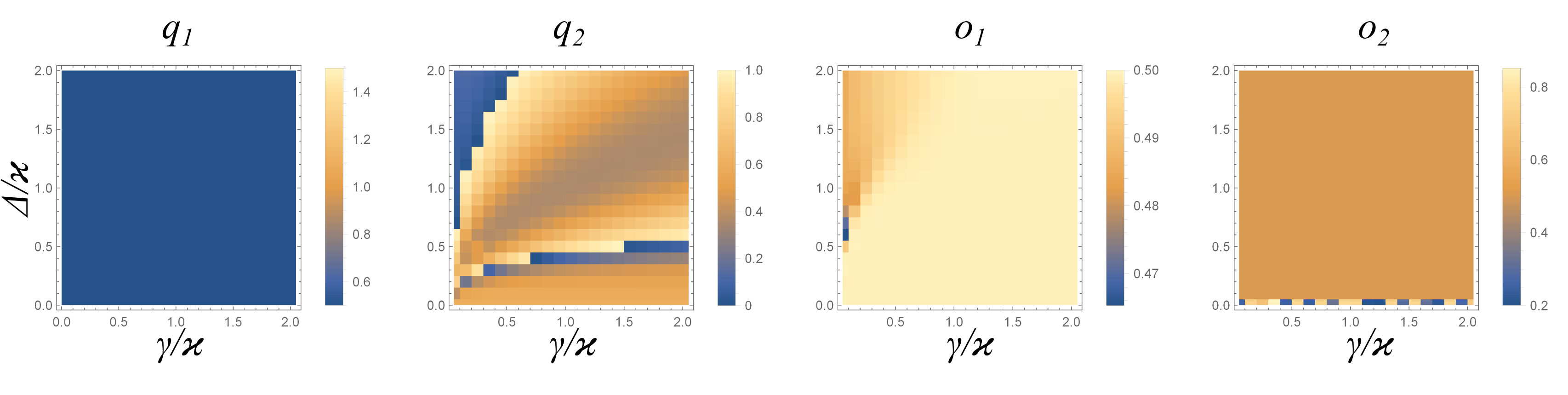}
	\caption{
		Numerically calculated quadrupole ($q_{1,2}$) and 2D-octupole ($o_{1,2}$) moment components for the two-mode Kagome lattice with various coupling constants $\gamma$ and $\Delta$. 
	}
	\label{Fig_moments}
\end{figure}

We then take the finite $108$-cite hexagonal sample of our two-mode kagome lattice, write the position operators in the basis of the Hamiltonian, and calculate the components of the quadrupole $(\hat{q}_1,\hat{q}_2)$ and 2D-octupole $(\hat{o}_1,\hat{o}_2)$ moments for open boundary conditions via Eq.~\eqref{real-space moments}, see Fig.~\ref{Fig_moments}. 
The calculation consistently shows that while $q_1 = 0.5$ is half-quantized, $q_2$ component changes continuously depending on the coupling constants. Thus, although our model possesses a finite quadrupole moment, one its component is not quantized and thus could not be linked to the strictly quantized corner charges and higher-order topology of our system. 
In sharp contrast, the calculated 2D-octupole moments $o_1$, $o_2$ are \textit{both} close to $0.5$ (half-quantized) for almost arbitrary coupling constants, as long as the bandgap is sufficient (while in the thermodynamic limit quantization happens for arbitrary $\gamma$ and $\Delta\neq 0$ as the bandgap is opened by infinitesimally small $\Delta$). 

\textit{Therefore, our model possesses a fully quantized 2D-octupole moment which could be regarded as a suitable topological invariant}. 
This feature differentiates our model from the canonical $C_2$- and $C_4$-symmetric quadrupole insulators which have non-quantized values of $o_1$ and $o_1$, and further proves that our model possesses higher-order multipole topology of a novel type. 
We thus can view the designed system as the first, to the best of our knowledge, example of a \textbf{two-dimensional octupolar HOTI}.



\section{Localization of corner states}

\begin{figure}[h!]
	\centering
	\includegraphics[width=0.25\textwidth]{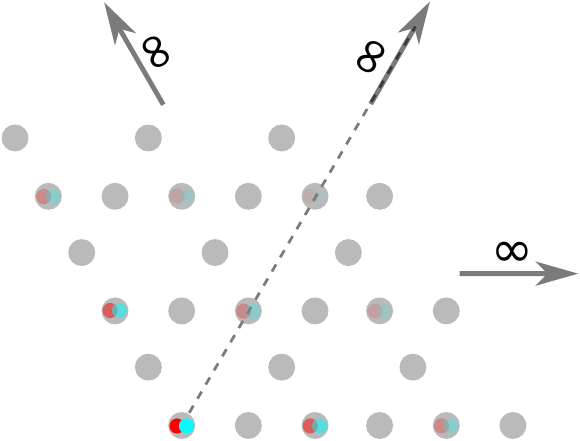}
	\caption{
		Corner state of the type encountered in the hexagon. 
	}
	\label{hexagon_corner_state}
\end{figure}

\begin{figure}[h!]
	\centering
	\includegraphics[width=0.35\textwidth]{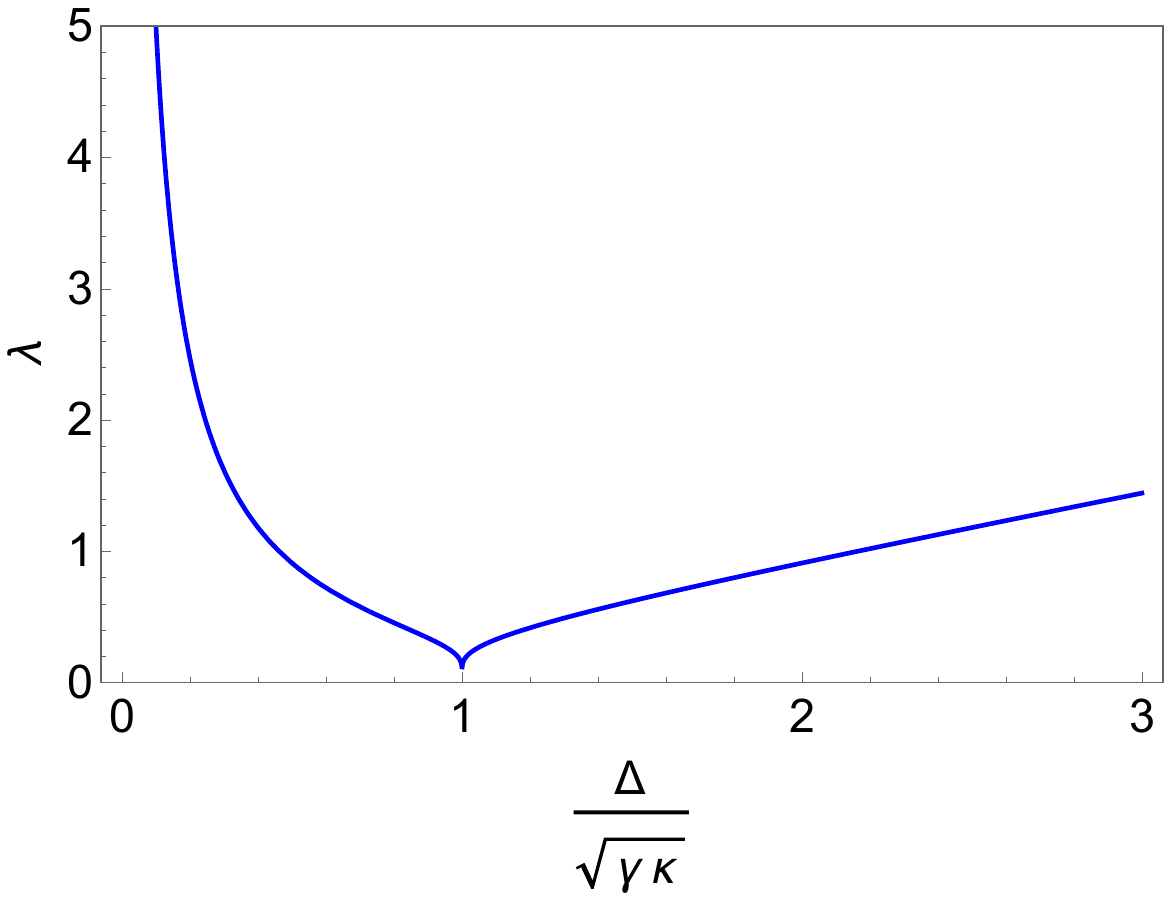}
	\caption{
		Localization length $\lambda$ as a function of the effective band splitting parameter $ \Delta / \sqrt{\gamma \kappa} $. Notice perfect localization $ \lambda = 0 $ at $\Delta = \sqrt{\gamma \kappa}$. 
	}
	\label{loc_length}
\end{figure}

Consider an isolated corner state at $2\pi/3$ angle in a semi-infinite geometry,  Fig.\ref{hexagon_corner_state}. 
We suppose that in the limit of an infinite obtuse lattice angle, the corner state is pinned to zero energy, and is localized only at one sublattice (see the inset in Fig.~2a in the main text). 
The coupled-mode equations then read 
\begin{eqnarray}
\label{12_hex}
&&\kappa_+ \textbf{u}_{m,n,1} + 
\kappa_- \textbf{u}_{m+1,n,1}
=
0
,\\
\label{13_hex}
&&\kappa_- \textbf{u}_{m,n,1} + \kappa_+ \textbf{u}_{m,n-1,1} 
=
0
, 
\end{eqnarray}
where $\textbf{u}_{m,n,i} \equiv \{s_{m,n,i},f_{m,n,i}\}$. Next, we choose the simplest anzatz describing uniform exponential decay from the corner, 
\begin{eqnarray}
\label{anzatz_hex}
\textbf{u}_{m,n,1} = \textbf{u}_{0,0,1} \cdot e^{-\alpha (n + m)}
.
\end{eqnarray}
Substituting \eqref{anzatz_hex} into Eqs.~\eqref{12_hex}-\eqref{13_hex}, we find that the latter lead to to the same matrix equation: 
\begin{eqnarray}
\label{meq}
\kappa_+ \textbf{u}_{0,0,1} + 
\kappa_- \textbf{u}_{0,0,1} \cdot e^{-\alpha}
=
0
.
\end{eqnarray}
At this point, we choose the solve-for vector in the form $ \textbf{u}_{0,0,1} = \{ 1, y \}^{T} $ (normalization factor is dropped). Equation~\eqref{meq} then has two solutions
\begin{eqnarray}
\label{meq_sols}
y = \pm i \sqrt{\kappa / \gamma}
, \qquad 
\alpha = \ln{\left[ \frac{2\Delta}{\Delta \mp \sqrt{\gamma \kappa}} - 1 \right]}
.
\end{eqnarray}
However, the second solution (with ``$-$'' in $ y $) is unphysical since it  exponentially grows away from the corner. Thus we are left with one corner-state solution: 
\begin{eqnarray}
\label{meq_sols}
\textbf{u}_{0,0,1} = \{ 1, i \sqrt{\kappa / \gamma} \}^{T}
, \qquad 
\alpha = \ln{\left[ \frac{2\Delta}{\Delta - \sqrt{\gamma \kappa}} - 1 \right]}
, 
\end{eqnarray}
which implies two cases with different phase profiles of the corner mode: when $ 0 < \Delta < \sqrt{\gamma \kappa} $, solution includes the factor $(-1)^{m+n}$ and the phase is alternating, and when $ \Delta > \sqrt{\gamma \kappa} $, the phase is homogeneous, which perfectly matches the phase profiles we have found numerically in the finite hexagon. 
In both cases, the localization length $\lambda$ is determined by Eq.~(2) in the main text, 
see Fig.~\ref{loc_length}. 

\section{Numerics for finite lattices with and without disorder}

\begin{figure}[H]
	\centering
	\includegraphics[width=0.50\textwidth]{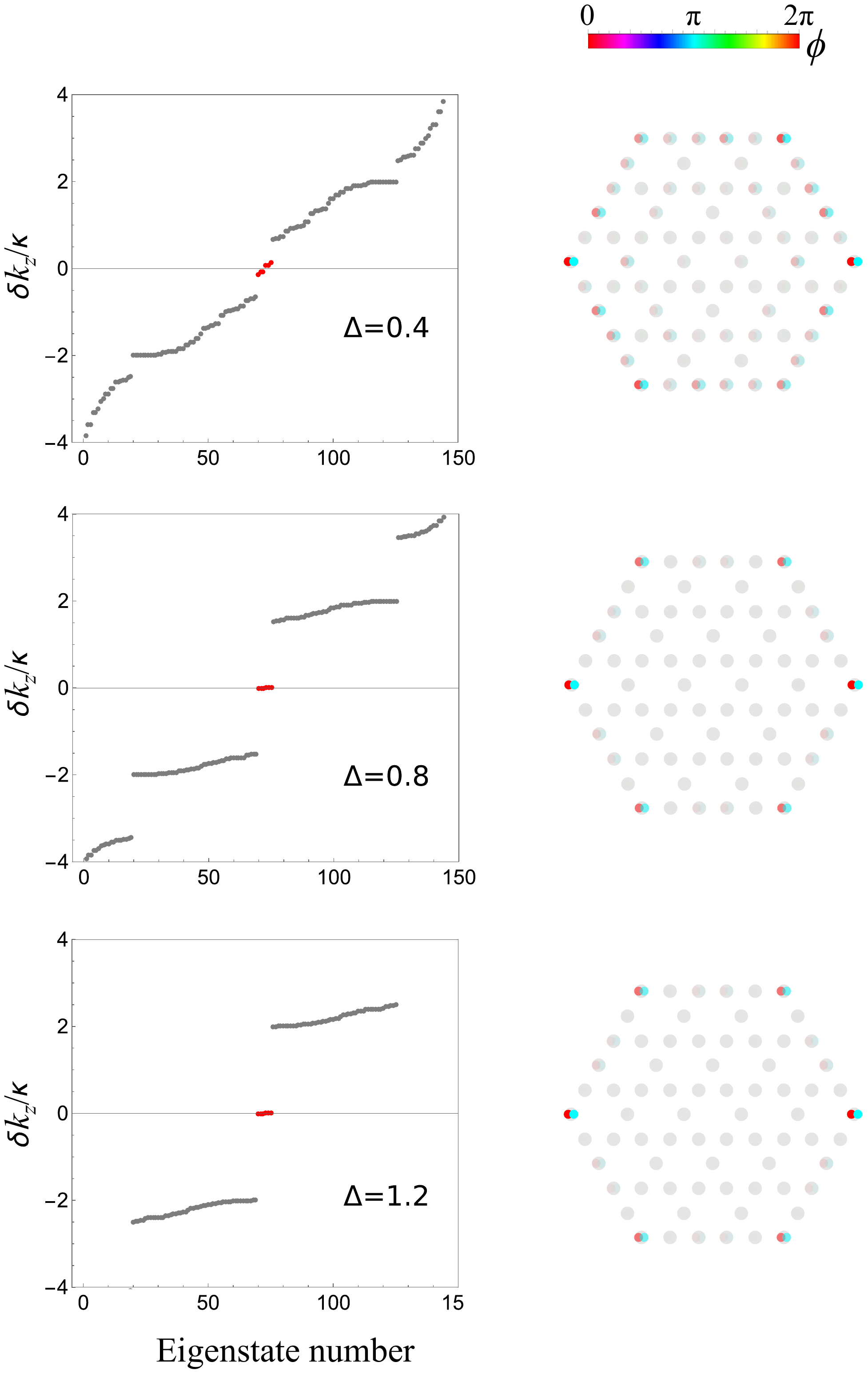}
	\caption{
		Corner states (marked by red) in a hexagonal sample (shown on right panels) for $ \kappa=1 $, $ \gamma=1 $ and different values of inter-orbital interaction $ \Delta $ (specified on plots). The phase profiles of the corner modes are color-coded. 
	}
	\label{colored_eigenstates}
\end{figure}

Examples of numerical spectrum of a hexagonal sample of the two-mode kagome lattice is shown in Fig.\ref{colored_eigenstates}. 
We see that in the topological region $\Delta\neq 0$ the lattice always features six corner states.



\begin{figure}[H]
	\centering
	\includegraphics[width=0.85\textwidth]{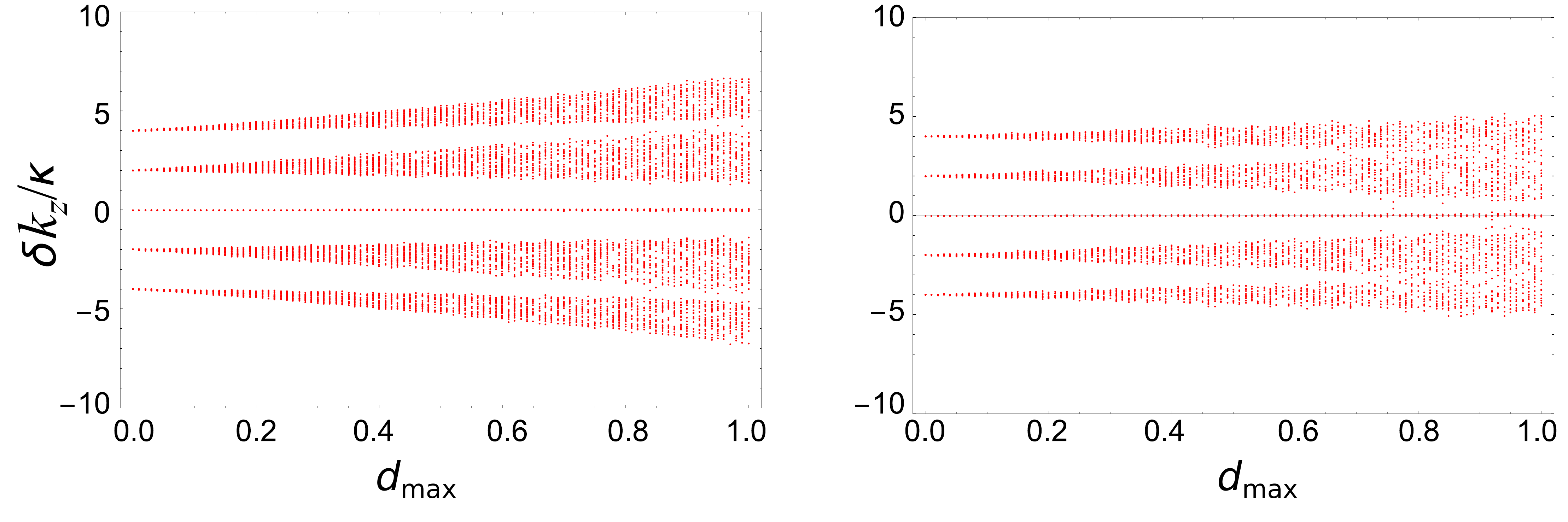}
	\caption{
		Evolution of the eigenstates of a finite hexagonal lattice of $135$ sites when introducing the disorder. 
		Left panel: allowing only positive disorder coefficients, $d_i \in \{0,\kappa \cdot d_{max}\},\, i \in \{1,2,3\}$
		Right panel: allowing for negative disorder coefficients, $d_i \in \{-\kappa \cdot d_{max},\kappa \cdot d_{max}\},\, i \in \{1,2,3\}$. 
		Parameters: $\kappa=\gamma=\Delta = 1$. 
	}
	\label{disorder1}
\end{figure}

Next, we introduce the disorder into the coupling matrices without breaking the Hermiticity of the full tight-binding Hamiltonian. For each coupling matrix, we introduce three possible types of disorder: 
\begin{eqnarray}
&&\hat{M}_{d1}
=
d_1 \cdot 
\left(
\begin{array}{cccccc}
1 & 0 \\
0 & 0 \\
\end{array}
\right)
, \\
&&\hat{M}_{d2}
=
d_2 \cdot 
\left(
\begin{array}{cccccc}
0 & 0 \\
0 & -1 \\
\end{array}
\right)
, \\
&&\hat{M}_{d3}
=
d_3 \cdot 
\left(
\begin{array}{cccccc}
0 & i \\
i & 0 \\
\end{array}
\right)
,
\end{eqnarray}
where $d_i \in \{0,\kappa \cdot d_{max}\},\, i \in \{1,2,3\}$ are three positive different disorder coefficients, and $d_{max}$ is a ``global'' maximum disorder constant normalized to $\kappa$ which is set for each consecutive eigenvalues calculation. 
The results for the hexagonal lattice are shown in Fig.\ref{disorder1}. We see that the zero-energy corner states typically retain their near-zero energy and are well-separated from the bulk states even for the large values of disorder $d_{max} \sim 1$. 

\section{Effect of imperfect mode degeneracy}

\begin{figure}[H]
	\centering
    \includegraphics[width=0.80\textwidth]{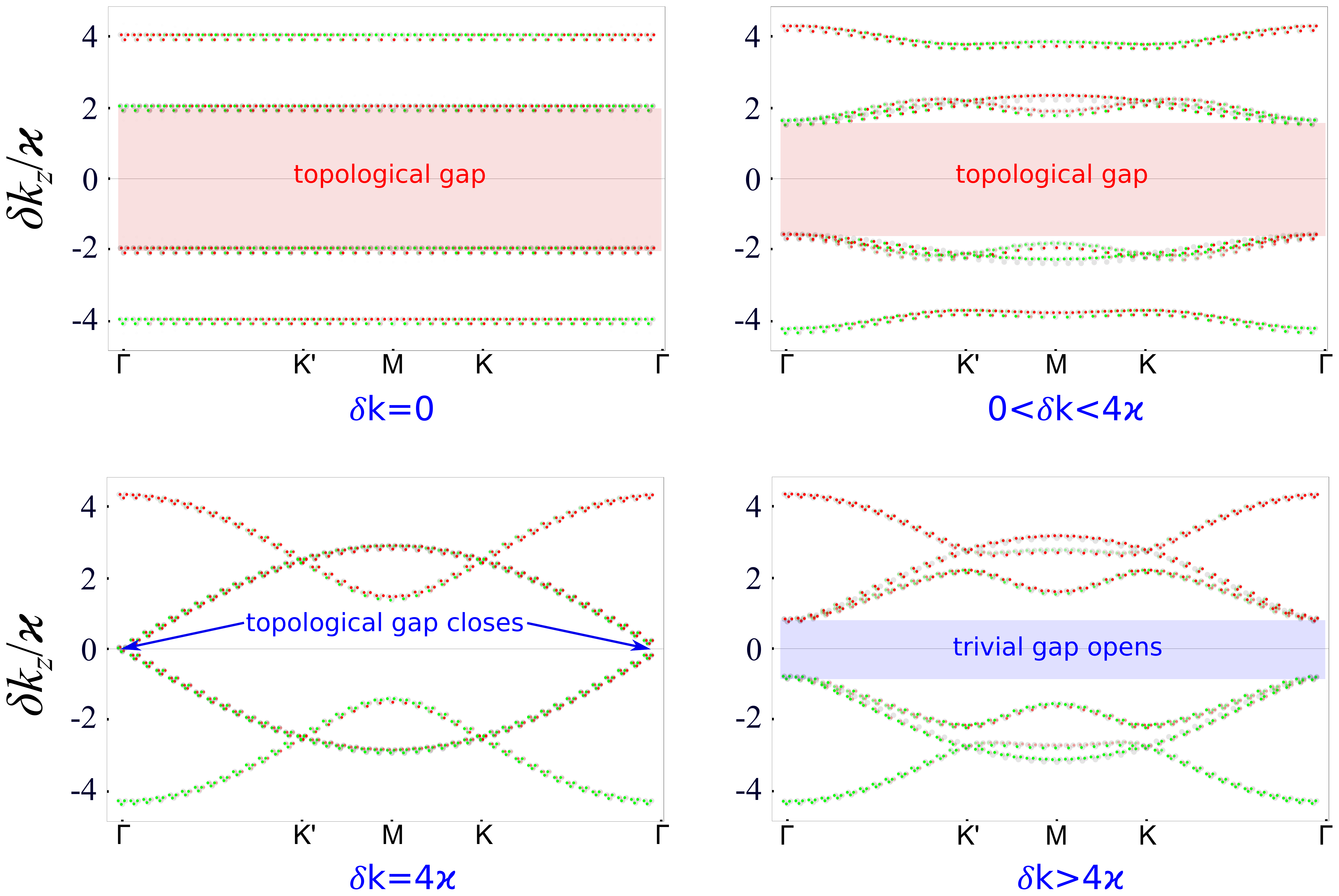}
    \caption{Band structure deformations with increasing wavevector detuning in each waveguide (shown here for symmetric case $\kappa=\gamma=\Delta$). }
	\label{gap_reopening}
\end{figure}

In this section, we discuss the effect of detuning of the modes in the waveguides, which could be caused by the imperfect waveguide design. 
We suppose that the individual waveguides are still identical, but the two modes of interest in each waveguide at a given frequency have slightly different propagation constants $ \delta k = k_{z1} - k_{z_2} $. In the Hamiltonian Eq.~\ref{Hamk}, this gives rise to the diagonal terms, $\hat{H}'(\textbf{k}) = \hat{H}(\textbf{k}) + \delta \hat{H}$, where $\delta \hat{H} = (\delta k / 2) \cdot \hat{\sigma}_z \otimes \hat{\sigma}_z \otimes \hat{\sigma}_z$, and $\hat{\sigma}_z = \{1,0;0,-1\}$ is the corresponding Pauli matrix. Using this perturbed Hamiltonian, we plot the example of band structure in Fig.\ref{gap_reopening} and observe the topological gap closing at $\Gamma$ point for a critical value of wavevector detuning $\delta k_{c}$; for $\delta k > \delta k_{c} $, the gap reopens, now being topologically trivial and hosting no corner states. 
This critical detuning therefore serves as a measure of maximum waveguide design imperfections which would not disturb the topological corner states.

\begin{figure}[H]
	\centering
    \includegraphics[width=0.90\textwidth]{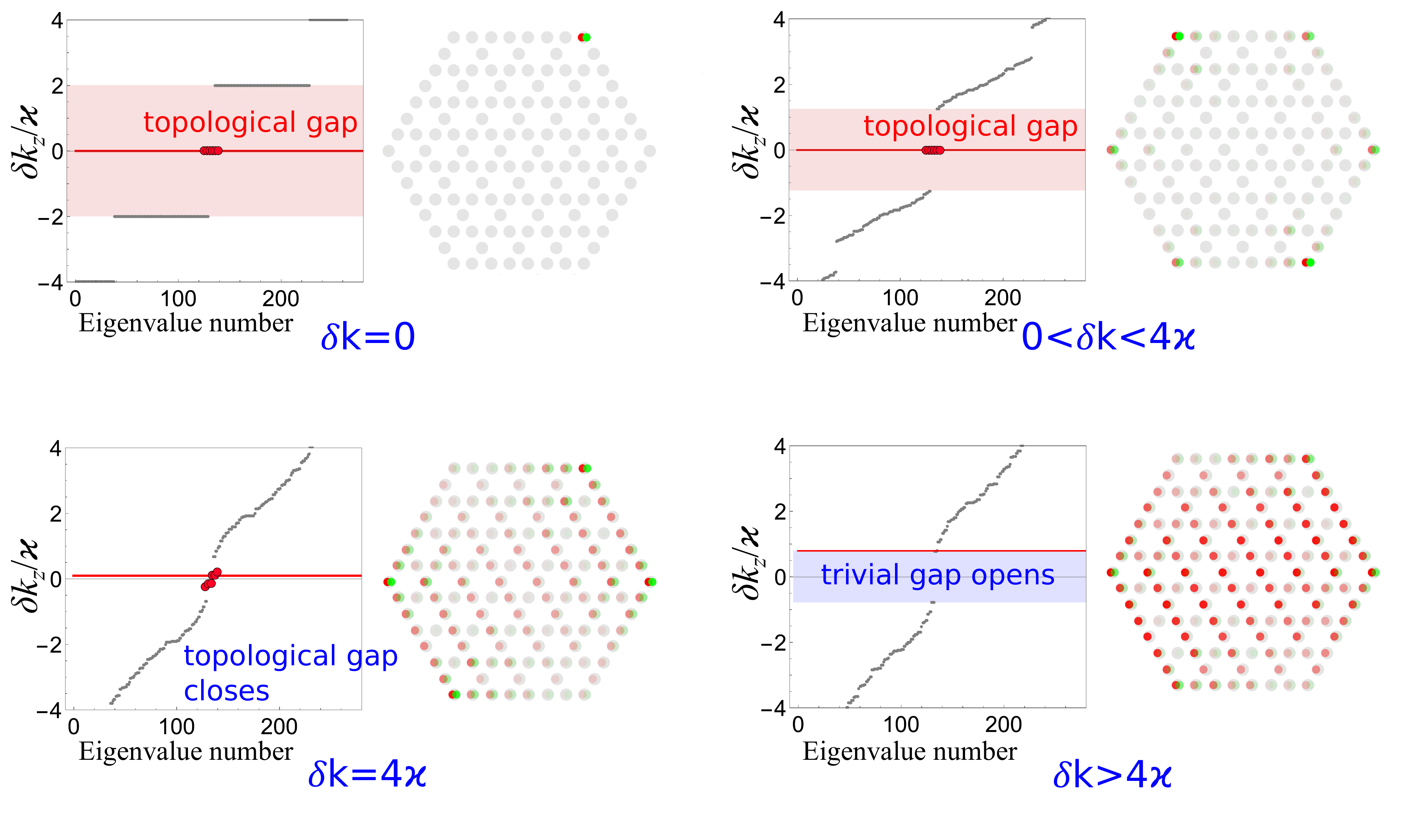}
    \caption{Numerics for the corner states stability with increasing wavevector detuning in each waveguide (for the case $\kappa=\gamma=\Delta$). One representative eigenstate is shown for each $\delta k$, with red horizontal line in the spectrum showing its energy. }
	\label{gap_reopening_numerics}
\end{figure}

The spectrum of $\hat{H}'(\textbf{k})$ at $\Gamma$ point in the general case reads $\{-4 \gamma -\delta k / 2,2 \gamma -\delta k / 2,2 \gamma -\delta k / 2,\delta k / 2-2 \kappa ,\delta k / 2-2 \kappa ,\delta k / 2+4 \kappa \}$. The critical detuning $\delta k_{c}$ therefore corresponds to the merging of two degenerate pairs of levels into four-times-degenerate group, which gives 
\begin{equation}
\delta k_{c} = 2( \gamma + \kappa )
. 
\end{equation}
This result is reproduced by numerical calculations summarized in Fig.\ref{gap_reopening_numerics}. We see that in the case $\kappa=\gamma$, the corner states survive and stay pinned to zero energy as long as the topological gap closes at critical detuning $\delta k_{c} = 4\kappa$; for larger $\delta k$ values, the corner states attach to continuum and disappear. For the general case $\kappa \neq \gamma$, the only difference is that at $\delta k < \delta k_{c}$ the corner states move to higher eigenvalue still staying perfectly sixfold degenerate. 
Note that for any parameters, no additional edge states appear, signalling that at nonzero detuning $\delta k \neq 0$ bulk polarization still remains zero.

\section{Experimental realization of the proposed design}

\begin{figure}[hb!]
	\centering
    \includegraphics[width=0.95\textwidth]{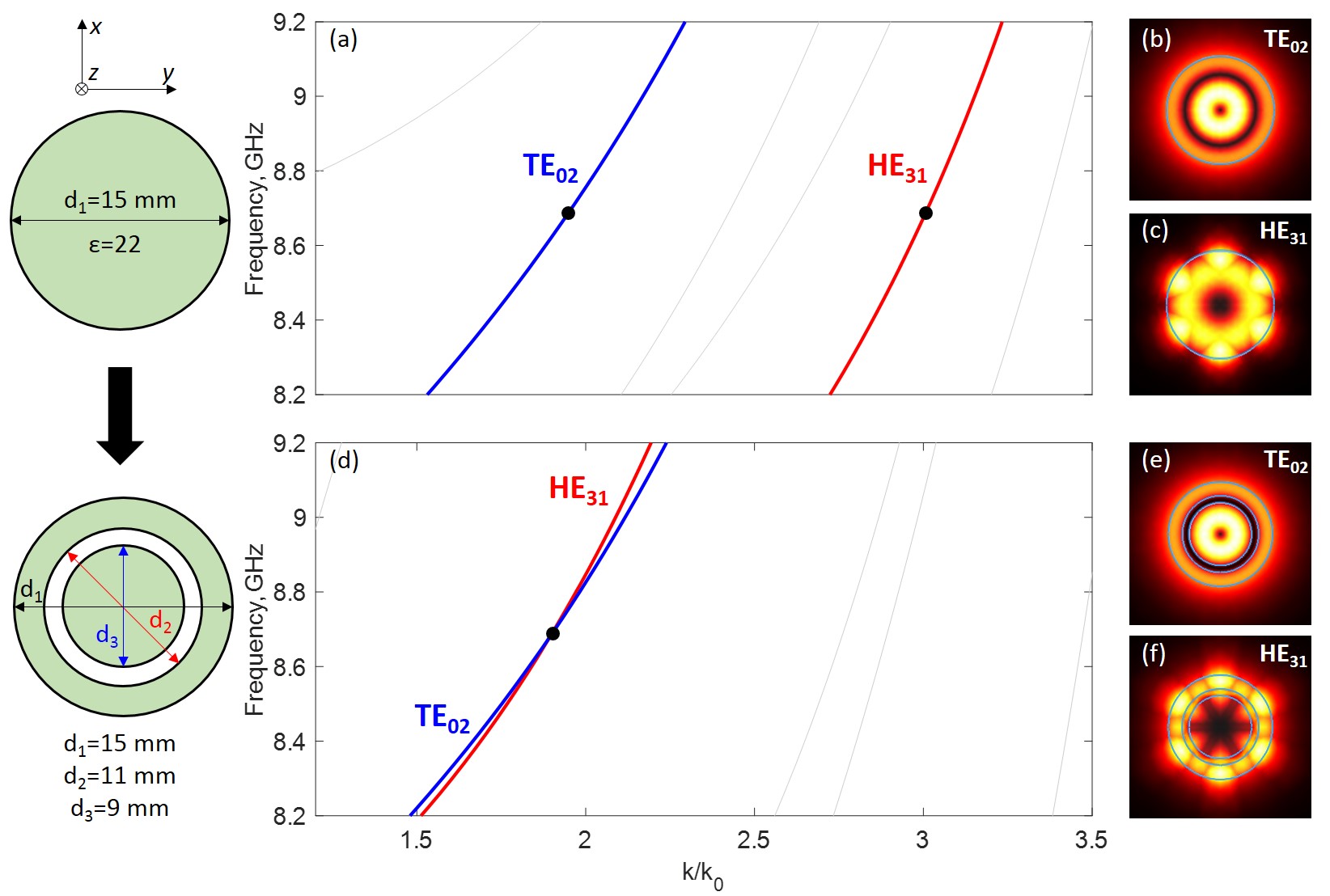}
    \caption{(a,d) Dispersion of the cylindrical waveguides with the cross sections shown on the left. Blue and red curves indicate TE$_{02}$ and HE$_{31}$ modes, respectively, which are employed in the design. (b,c,e,f) Electric field distribution $|\mathbf{E}|$ for (b,e) TE$_{02}$ and (c,f) HE$_{31}$ modes, respectively; (b,c) correspond to the homogeneous waveguide, (e,f) correspond to the waveguide with the air gap.}
	\label{fig:singleWG_design}
\end{figure}

The main building block of the proposed topological lattice~-- a waveguide with two quasi-degenerate modes with azimuthal numbers 0 and 3~-- was realized by introducing the air gap in the initially homogeneous cylindrical ceramic waveguide with permittivity $\eps=22$, as shown in Fig.~\ref{fig:singleWG_design}. The dispersion of the homogeneous waveguide with diameter 15~mm is shown in Fig.~\ref{fig:singleWG_design}(a). The chosen modes TE$_{02}$ and HE$_{31}$ are well separated in reciprocal space with the electric field distributions at the frequency $8.7$~GHz shown in Figs.~\ref{fig:singleWG_design}(b,c). The mode TE$_{02}$ possesses a minimum of electric field at the radial distance $r \approx 5$~mm. Introducing a ring-like air gap at this radial position, therefore, only slightly changes the dispersion of this mode. The electric field of the mode HE$_{31}$, on the other hand, is rather strong at this position, and, therefore, its dispersion significantly depends on the thickness of the gap. For the thickness 1~mm the dispersion of the TE$_{02}$ and HE$_{31}$ modes intersect at the frequency $\approx$8.7~GHz.

\begin{figure}[ht!]
	\centering
    \includegraphics[width=0.80\textwidth]{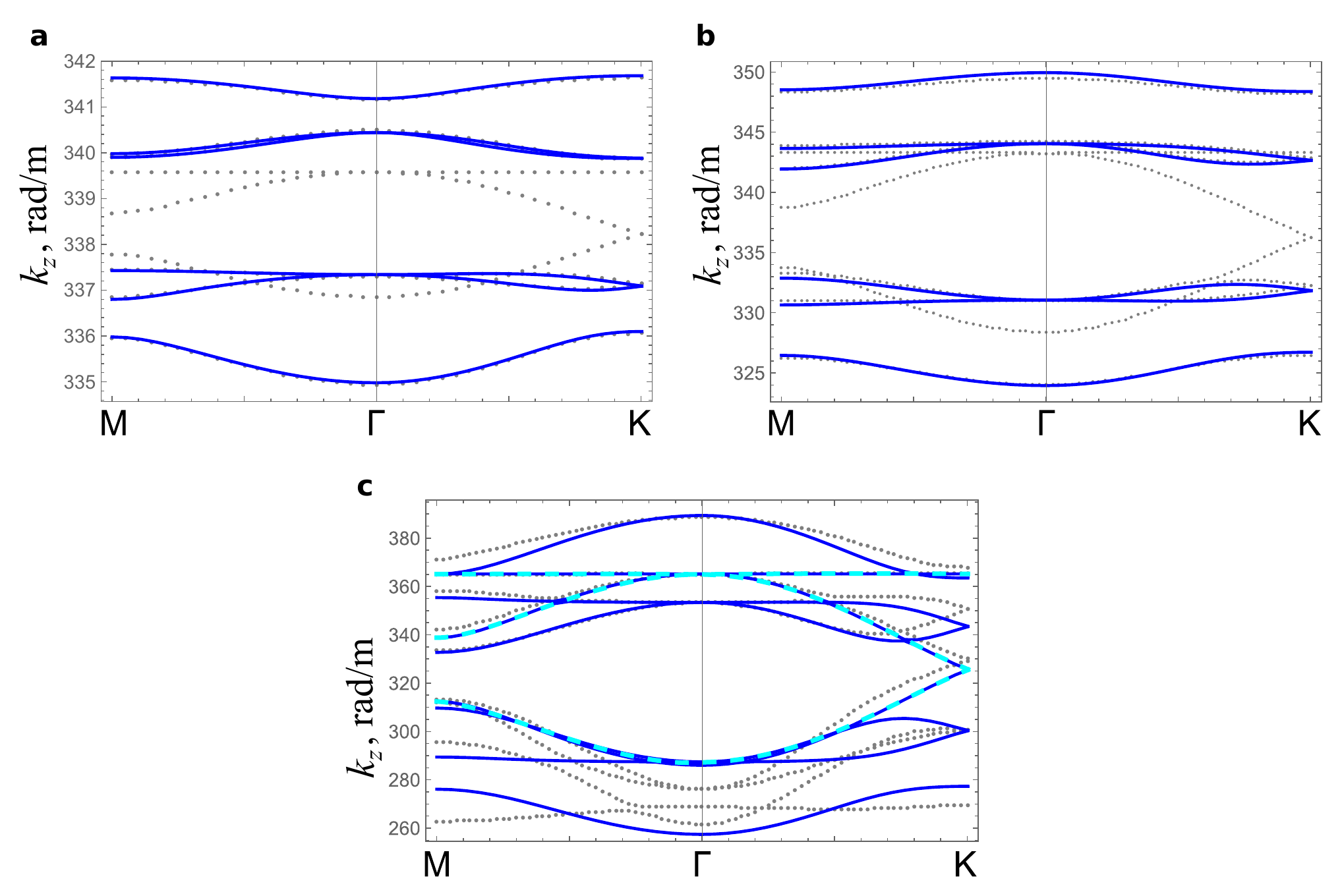}
    \caption{Numerical calculated dispersion $k_z(k_{||})$ along two high-symmetry directions $\text{M}-\Gamma-\text{K}$ (gray dots) and corresponding best analytical fits (solid blue curves) for three different periods: \textbf{a} $25$mm; \textbf{b} $21$mm; \textbf{c} $17$mm (as in experiment). Parameters used for numerical fits (all in rad/m units): 
    \textbf{a}, $\kappa=0.64, \, \gamma=0.91, \, \Delta=0.67, \, k_{z0}=338.62 $; 
    \textbf{b}, $\kappa=3.35, \, \gamma=3.15, \, \Delta=2.5, \, k_{z0}=337.35 $; 
    \textbf{c}, $\kappa=17, \, \gamma=16, \, \Delta=8, \, k_{z0}=321.4 $. In panel \textbf{c}, dashed cyan lines mark the analytical spectrum of rotated $f$-modes with parameters $k'_{z0}=338.8$~rad/m and $\gamma'=13.2$~rad/m. 
    }
	\label{fig:lattice_design}
\end{figure}

\begin{figure}[hb!]
	\centering
    \includegraphics[width=0.72\textwidth]{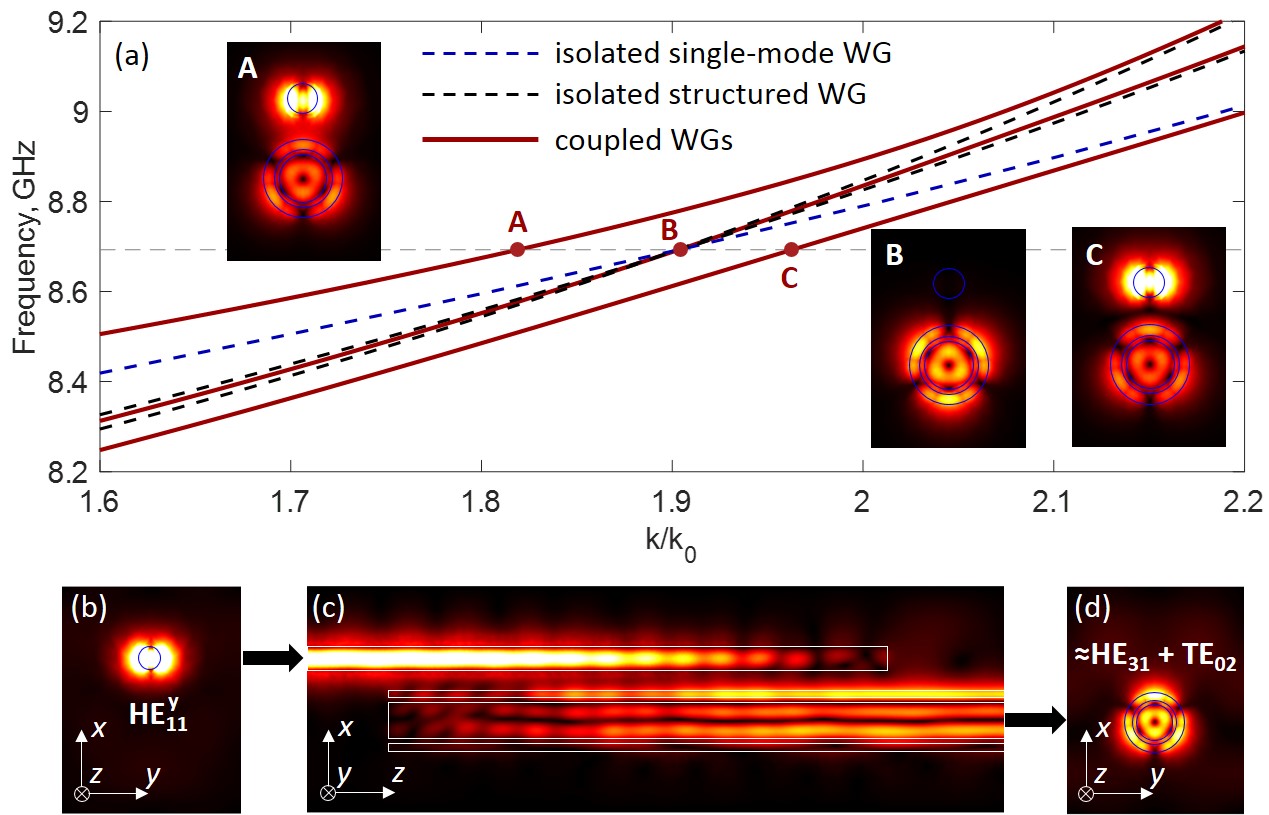}
    \caption{(a) Dispersion of the isolated multimode (dashed black curves) and singlemode (dashed blue curve) waveguides and coupled waveguides (solid red curves); only modes with vertical $x-z$ electric symmetry plane are shown. Insets show electric field distribution $|\mathbf{E}|$. (b) Fundamental mode of the single-mode waveguide transforms via directional coupler (c) to the superposition of TE$_{02}$ and HE$_{31}$ modes of the multimode waveguide (d).}
	\label{fig:DC}
\end{figure}

The dependence $k_z(k_{||})$ of the kagome lattice of the designed multimode waveguides for the frequency 8.65~GHz is shown in Fig.~\ref{fig:lattice_design} for the three values of the period $a$. We observe that for relatively large periods $a=24$~mm and $a=21$~mm, Fig.~\ref{fig:lattice_design}(a-b), the numerically calculated band structure shown with grey dots is well described within the tight-binding approximation shown with blue solid curves. In the experiment, we choose a smaller value of the period $a=17$~mm in order to measure the discrete diffraction of the bulk states using rather short waveguides. Although for such small periods the picture becomes more complicated due to the next-nearest neighbour couplings and inaccuracy of the tight-binding approximation, the qualitative picture remains the same, as shown in Fig.~\ref{fig:lattice_design}(c). Despite the imperfect fit by the analytical model and emergence of the additional band hybridization, the bulk gap for the original $s$ and $f$ modes remains open and well described by the tight-binding model. Moreover, the fact that the bandgap remains open for decreasing lattice periods suggests that all three cases share the same higher-order topology. In particular, calculations of the finite size array of 30 waveguides confirm the presence of 6-fold degenerate corner states with high degree of localization (see Fig.~3 in the main text).

Excitation of the waveguide array in the experiment was performed via the extended corner waveguide. Switching between the excitation of the corner and bulk states of the array was accomplished by tuning a proper superposition of the waveguide modes in the extended waveguide, which, in turn, was achieved by its directional coupling with an auxiliary single-mode waveguide (SMWG). In Fig.~\ref{fig:DC}(a) we show the dispersion of the isolated (dashed curves) and coupled (solid curves) cylindrical multimode waveguide and SMWG; the SMWG is placed above the multimode one. The diameter of the SMWG $d=8.5$~mm was chosen in such a way that the dispersion of its fundamental mode HE$_{11}$ intersects the crossing point of TE$_{02}$ and HE$_{31}$ modes in the multimode waveguide. As one can observe from the field distributions of the coupled modes at the frequency $\approx 8.7$~GHz (points ``A'', ``B'', ``C''), shown on the insets in Fig.~\ref{fig:DC}(a), one of the coupled modes (point ``B'') represents a superposition (HE$_{31}$ - TE$_{02}$) of two modes of the multimode waveguide, which is uncoupled from the SMWG. Consequently, the HE$_{11}^y$ mode (with dominant $y$ component of the electric field) of the SMWG couples only to the orthogonal superposition (HE$_{31}$ + TE$_{02}$) in the given arrangement, as indicated by the insets ``A'' and ``C''. Calculation of the wave propagation in the directional coupler excited with the HE$_{11}^y$ mode of the SMWG [Fig.~\ref{fig:DC}(b-d)] shows that $\approx$ 98\% of the power is transferred to the multimode waveguide at the operational frequency. By changing the position of the SMWG to the opposite with respect to the $x$ axis, one changes the superposition of the excited modes from (HE$_{31}$ + TE$_{02}$) to (HE$_{31}$ - TE$_{02}$), thus allowing to switch between the excitation of corner and bulk modes of the array.

\begin{figure*}[ht!]
	\centering
	\includegraphics[width=0.85\textwidth]{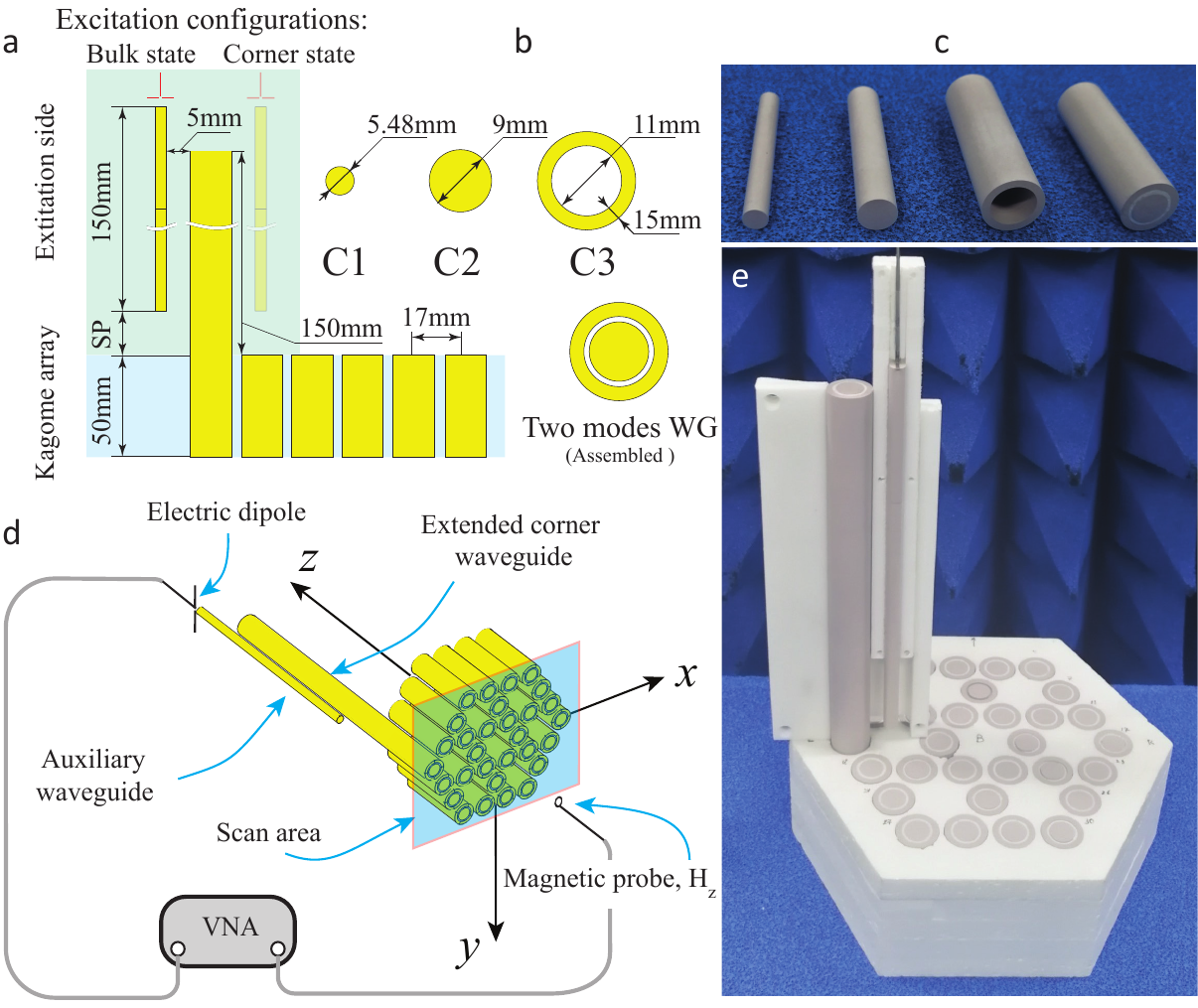}
	\caption{
		(a) The schematic side-view of the experimental sample with geometry parameters and excitation configurations. 
		(b) Side view of the cylinders used for the fabrication of the array, shown in (c), from left to right: auxiliary waveguide C1; core cylinder C2 of the two-mode waveguide; outer hollow cylinder C3 of the two-mode waveguide; assembled two-mode waveguide. 
		(d) The excitation scheme. 
		(e) Side view of the fabricated waveguide array in an anechoic chamber. 
	}
	\label{FigExpS}
\end{figure*}

Further details of the experimental setup are summarized in Figs.~\ref{FigExpS}
(see also the Methods section).

